\documentclass[10pt]{article} % For LaTeX2e
\usepackage[preprint]{tmlr}

\usepackage{amsmath,amsfonts,bm}

\def\eqref#1{equation~\ref{#1}}
\def\1{\bm{1}}

\DeclareMathAlphabet{\mathsfit}{\encodingdefault}{\sfdefault}{m}{sl}
\SetMathAlphabet{\mathsfit}{bold}{\encodingdefault}{\sfdefault}{bx}{n}

\usepackage{hyperref} 
\usepackage{subcaption}
\usepackage{url}
\usepackage{booktabs} 
\usepackage{graphicx,enumitem}
\usepackage{amsmath,amssymb,amsfonts}
\usepackage{amsthm}
 \newtheorem{theorem}{Theorem}

\usepackage[export]{adjustbox}
\theoremstyle{definition} 
\newtheorem{corollary}{Corollary}

\usepackage{tcolorbox}
\tcbuselibrary{listings,breakable}

\newtcblisting{promptbox}{
    listing only,
    breakable,
    colback=black!2,
    colframe=black!25,
    boxrule=0.35pt,
    arc=1mm,
    left=1mm,
    right=1mm,
    top=1mm,
    bottom=1mm,
    listing options={
        basicstyle=\ttfamily\footnotesize,
        breaklines=true,
        columns=fullflexible,
        keepspaces=true
    }
}

\theoremstyle{remark}

\newcommand{\dtime}{t}

\title{Emergence of Preferential Attachment and Glass-Ceiling Effects in  Autonomous Networks of LLMs\thanks{This work was supported by National Science Foundation  grant   CCF-2312198  and Army Research Office grant W911NF-24-1-0083.}}

\author{\name Yiming Zhang \email yz2926@cornell.edu \\
      \addr Department of Electrical and Computer Engineering\\
      Cornell University
      \AND
      \name Vikram Krishnamurthy \email vk342@cornell.edu \\
      \addr Department of Electrical and Computer Engineering\\
      Cornell University}

\def\month{05}  % Insert correct month for camera-ready version
\def\year{2026} % Insert correct year for camera-ready version
\def\openreview{\url{https://openreview.net/forum?id=XXXX}} % Insert correct link to OpenReview for camera-ready version

\begin{document}

\maketitle

\begin{abstract}
We investigate the emergence of structural disparities in networks comprising large language model (LLM) agents. 
Each LLM agent refers to a prompted LLM of a specified type determined by its base model, model size, and system prompt.
When LLM agents autonomously choose collaborators, the resulting communication network exhibits preferential-attachment dynamics: agent types that are already prominent become increasingly likely to attract additional connections. 
In some cases, weaker LLM agents 
(agents with smaller base model or older version) can disproportionately occupy central and influential network positions relative to stronger LLM agents. We interpret this  misalignment between task capability and network prominence as a type-dependent \textit{glass-ceiling effect} (GCE).

We model the network of LLM agents as a time-evolving sequence of directed weighted graphs, where the vector-valued edge weights represent cumulative tokens exchanged, number of interaction rounds, and reasoning effort. 
Using a contraction mapping argument on the mean-field dynamics, we prove that the importance (centrality) of each agent type converges to a unique stable equilibrium. 
To anchor the model in LLM decision mechanisms, we introduce a cross-attention-inspired utility for collaborator selection.
This utility specifies the local connection dynamics and, together with the mean-field model, yields a predictive characterization of the limiting network structure and its type-dependent centrality gaps.

To validate the theory, we develop an experimental testbed with 100 LLM agents. 
Our experiments show that autonomous network formation can generate persistent centrality disparities, with their magnitude and direction depending on model family, model size, system-prompt design, and task context. 
They further show that the effect of preferential attachment depends on its alignment with model capability: reinforcing it improves collective performance when stronger agents become central, whereas weakening it improves performance when network dynamics instead favor weaker agents.
All results are reproducible; the code and datasets are available in an  \href{https://anonymous.4open.science/r/LLM_GCE-36F3/}{anonymous GitHub repository}.
\end{abstract}

% \vspace{-5mm}
\section{Introduction}
% \vspace{-3mm}

Multi-agent LLM networks, where each node is an LLM agent\footnote{Throughout this paper, an \emph{LLM agent} refers to  a prompted LLM of a specified type. An agent's type is determined by its  base model (e.g., Gemini or GPT), model size (number of parameters), and system prompt defining its role (e.g., answer provider or answer checker).} are becoming increasingly important as large language models are deployed not only as isolated assistants, but also as interacting agents capable of collaboration, specialization, and collective problem solving.
Such networks of LLM agents  have been explored in a range of settings, including complex problem solving \citep{qian2024scaling}, 
software development \citep{qian2024chatdev,hong2024metagpt}, 
automated debate and collective judgment \citep{li2024sparse,hu2025multiagent}, 
and large-scale social simulation \citep{piao2025agentsociety,guan2025modeling}.
As LLMs continue to improve through scaling and instruction tuning, they are increasingly studied not only as problem-solving tools, but also as agents that can exhibit communicative behaviors, preferences, biases, and social interactions \citep{park2023generative,ashery2025emergent,madmoun2025communication}.
Recent work examines multi-agent LLM systems from a social-network perspective, where agents exchange information, influence one another, and form structured patterns of interaction \citep{papachristou2025network,mehdizadeh2025homophily,jain2025information,schneider2025learning}.
This perspective suggests several potential applications, such as simulating human social dynamics \citep{park2023generative,piao2025agentsociety,guan2025modeling}, supporting collective decision-making in organizations \citep{qian2024scaling,guo2026coalition}.
These applications require understanding how large-scale LLM agent networks emerge when agents are allowed to interact and form connections.

% \newpage

As LLM agents autonomously form and evolve their own interaction networks, a central question arises: {\em Do complex sociological phenomena, including structural disparities observed in human social networks, also emerge in networks of LLM agents?} One important sociological phenomenon observed in social networks is preferential attachment  \citep{barabasi1999emergence}, a reinforcement mechanism in which agents that already have more connections or greater prominence become increasingly likely to attract additional connections. Here, an agent's centrality measures its prominence in the communication network, namely its access to and influence over interactions with other agents.
We distinguish between two types of preferential attachment: 

\vspace{-8pt}\begin{itemize}
\setlength{\topsep}{0pt}
\setlength{\partopsep}{0pt}
\setlength{\itemsep}{0pt}
\setlength{\parsep}{0pt}
\setlength{\parskip}{0pt}
    \item \textit{Capability-aligned dominance}: stronger agents\footnote{
    We use ``stronger'' and ``weaker'' to distinguish LLMs by their general model capability, as reflected in model version and number of parameters. A newer model version or a larger model is termed stronger than an earlier version or a smaller model.}   become central; this shows efficient specialization or merit-aligned preferential attachment. We term this as the \textit{meritocracy case}. This can improve system efficiency by encouraging stable specialization and reducing redundant interactions.
    \item \textit{Capability-misaligned dominance}: weaker or equal agents become central, or stronger agents remain peripheral; we term this as the \textit{glass-ceiling effect (GCE)}. In the social sciences, the GCE refers to invisible barriers that prevent certain groups of people from reaching higher organizational positions despite possessing comparable qualifications.  In networks of LLM agents, this can
     undermine collaboration by excluding
capable agents from important communication channels.
\end{itemize}

\vspace{-8pt}
%This paper combines tools from network science with extensive experiments to demonstrate the emergence such of preferential attachment in LLM agent networks.

This paper shows that LLM agents autonomously form networks that have a preferential attachment structure. Depending on the application, we show that  capability-aligned dominance or  
capability-misaligned dominance (GCE) can emerge. 
%This paper shows that an analogous glass-ceiling phenomenon emerges in autonomous multi-agent LLM networks: agents assigned to certain types may become structurally constrained in their access to high-value collaborations, even when their task capabilities are comparable to those of more central agents.
%Unlike structural inequality in human society, however, such disparities in LLM agent networks are not intrinsically undesirable or unfair.
This dual nature makes it important to characterize when type-dependent centrality gaps should be mitigated and when they can be intentionally leveraged for better network design.

\vspace{-3mm}
\subsection{Main Results and Insights}
\vspace{-2mm}
\paragraph{Autonomous network formation by LLM agents.}
First, we examine how LLM agents autonomously form interaction
networks while solving problem-solving tasks. In our framework, agents are not
instructed to follow any particular network-formation rule; instead, they
exchange information, assess the usefulness of received outputs, and decide
whether to establish or strengthen directed connections with other agents.
To analyze the resulting network evolution, we model the type-level
communication dynamics through a mean-field ordinary differential
equation (ODE), whose drift is derived from a cross-attention-based pairwise utility
function that approximates how individual LLM agents evaluate potential
collaborators and make connection decisions. This utility combines the semantic
relevance of a prospective collaborator's context with its structural
attractiveness in the current network, thereby capturing both the
informational value of the collaborator and its accumulated communication
position. 
We show that aggregating these local LLM decisions gives rise
to persistent communication patterns. The resulting mean-field
dynamics converge to a stable fixed point, providing a tractable characterization
of the long-run type-level communication structure.
\vspace{-8mm}
\paragraph{Emergence of preferential attachment and GCE.}
Second, we characterize when the 
%unequal network centrality structure induced by the 
autonomous network-formation protocol manifests as type-dependent preferential attachment. 
We study this question in two tasks: \textit{collaborative question-answering}, in which LLM agents exchange partial or noisy evidence to jointly answer a question, and \textit{multi-agent debate}, in which agents selectively inspect, challenge, and revise another’s arguments before producing final answers. 
Within each task, the  communication dominance pattern depends on agents’ model size, model family, and role-defining system prompts: communication prominence may be capability-aligned, with stronger agents becoming more central, or capability-misaligned, yielding a glass-ceiling effect (GCE). 
We quantify this dominance using type-level communication prominence and formalize average GCE, which captures population-wide centrality disadvantage, and tail GCE, which captures exclusion of agents from the most influential positions.
\vspace{-4mm}
\paragraph{Extensive experimental evaluation.}
We conduct four detailed experiments with networks evolved over
\(100\) LLM agent interactions.  
\textbf{(i)}
We demonstrate that 
%model LLM network-formation dynamics using 
a mean-field differential equation,  parameterized by a novel  \textit{cross-attention-based} pairwise utility predictor learned from LLM-generated connection decisions,  can accurately predict the evolution of network formation across different LLM agent contexts.
\textbf{(ii)} We examine how different forms of LLM heterogeneity shape preferential attachment. 
In capability-aligned settings, we show that stronger 
% (more parameters) 
same-family LLMs and closed-source 
LLM agents preferentially attract communication links. i.e., a meritocracy emerges.
In capability-misaligned settings, we show that prompt-defined roles result in weaker LLM agents taking on higher positions of  importance, and more powerful LLM agents are relegated to lower levels of importance, i.e., a glass-ceiling effect emerges.
\textbf{(iii)} After network formation, 
%we study how the capability-alignment between communication centrality and agent capability shapes information propagation: 
we demonstrate that truthful evidence is propagated in capability-aligned LLM networks, whereas hallucinations are amplified in capability-misaligned LLM networks.
\textbf{(iv)} Finally, by tuning a bias coefficient, we demonstrate improvements in network-wide answer accuracy in  question-answer  tasks: in capability-aligned settings, stronger preferential attachment is beneficial, while in capability-misaligned settings, weakening preferential attachment mitigates the GCE  and improves the quality of the agents' final outputs.

\vspace{-2mm}

\subsection{Motivation and Related Work}
\vspace{-3mm}

\textit{Multi-agent LLM systems and emergent social networks.} As LLMs are increasingly deployed for complex tasks, a growing body of work organizes them as interacting agents that communicate, divide tasks, critique outputs, and coordinate toward shared objectives \citep{guo2024large,wu2023autogen,li2023camel,chen2024agentverse}. Representative systems include AutoGen for flexible agent conversations \citep{wu2023autogen}, CAMEL and AgentVerse for type-based collaboration and emergent behaviors \citep{li2023camel,chen2024agentverse}, MetaGPT and ChatDev for structured software-development workflows \citep{hong2024metagpt,qian2024chatdev}, and multi-agent debate frameworks that improve reasoning through mutual critique \citep{du2024multiagentdebate}. However, these systems often rely on predefined roles, fixed workflows, or centrally specified communication protocols. A complementary line of work views LLM agents as social entities: generative agents exhibit individual and group behavior \citep{park2023generative}, while other studies use LLMs to simulate social and economic interactions \citep{aher2023using,argyle2023out,horton2023large}. Recent work further shows that decentralized LLM populations can develop social conventions, collective biases, and network structures such as hubs, communities, homophily, and preferential attachment \citep{gao2023s3,jain2024interacting,ashery2025emergent,papachristou2025network}. These findings motivate treating multi-agent LLM systems as artificial societies in which agents repeatedly choose collaborators and form persistent communication patterns. Our work builds on this perspective by studying how such communication networks emerge when LLM agents autonomously form and reinforce connections during collective problem solving.

\textit{Preferential attachment, structural inequality, and glass-ceiling effect.}
Our analysis is connected to classical work on cumulative advantage, preferential attachment, and structural inequality.
The Matthew effect explains how early success can reinforce future success \citep{merton1968matthew,price1976general}, while preferential attachment formalizes how well-connected nodes attract more links and produce persistent centrality differences \citep{barabasi1999emergence}.
Social-network theory further shows that network position shapes access to information, influence, and opportunities \citep{granovetter1973strength,burt1992structural}.
Recent work connects preferential attachment to glass-ceiling effect in directed or attributed networks, showing how homophily, group size, and cumulative advantage can generate asymmetric access to high-degree positions and structural disparities \citep{nettasinghe2022scale,nettasinghe2026emergence,luo2024mutual}.
The GCE  describes persistent barriers that prevent disadvantaged groups from reaching top positions \citep{cotter2001glass}.

\vspace{-4mm}
\section{Problem-Solving Tasks and Autonomous Collaboration-Network Formation}
\vspace{-2mm}
\label{sec:2}
In this section, we first describe the problem-solving tasks through which LLM
agents autonomously form collaborative networks. We then specify the protocol
rules governing how agents communicate, assess received outputs, and establish
connections with other agents. 
Because these interactions require
agents to assign connection weights based on the value of received information,
the protocol allows us to study how LLM networks form, evolve, and develop
type-level centrality gaps. This setting provides the foundation for the
GCE analysis in Sec.~\ref{sec:3} and the experiments in Sec.~\ref{sec:4}.

\subsection{Collaborative question-answering and Multi-Agent Debate}
\label{subsec:dataset_intro}

\textit{Task settings.}
We evaluate LLM network formation on two downstream tasks: collaborative question-answering and multi-agent debate.
Both tasks require agents to exchange information and choose whom to consult, making them suitable for studying autonomous connection formation.
In collaborative QA, agents observe partial or noisy evidence and collaborate to answer a question.
In multi-agent debate, agents first produce independent answers and arguments, then selectively inspect, challenge, or revise others' arguments before finalizing their answers.
For both tasks, we consider two agent types, $R$ and $B$, which may represent different base models or assigned roles.
This design allows us to test whether type-dependent centrality gaps emerge even when the two groups have similar competence.

\textit{CollaborativeQA and Multi-agent Debate Datasets.}
Similar to~\citep{jain2025collaborative}, we construct a synthetic dataset for both collaborative QA and multi-agent debate, with  implementation details provided in Appendix~\ref{app:experiment-details}. 
Here, the dataset refers to the collection of task prompts, agent-specific inputs, and the resulting LLM interaction traces and responses generated during the experiments.
We use this controlled construction rather than off-the-shelf benchmarks because it allows us to precisely specify agent capability, information access, type-level heterogeneity, and ground-truth outcomes. 
This control is essential for isolating how network structure emerges from agent interactions rather than from uncontrolled biases in pre-existing datasets.

For collaborative QA, each instance contains a question, a ground-truth answer, supporting evidence, and distractor snippets; each agent receives only a subset of the evidence, creating the need to consult others for missing information.
For multi-agent debate, each instance contains a question, the correct answer, plausible incorrect answers, and supporting or opposing arguments for each candidate; agents first generate independent answers with rationales, then selectively inspect or challenge others' arguments before revising their answers.
In both tasks, the two agent types are balanced to have comparable initial correctness, while their local evidence, arguments, confidence, or type prompts may vary.
Because the ground truth is known, we can measure whether interactions improve answer quality, whether useful evidence or high-quality arguments receive attention, and whether one type is structurally excluded from valuable collaborations despite comparable ability.
This enables analysis of influence, attention centrality, and GCE across both QA and debate networks.

\label{sec:model}
\vspace{-2mm}

\subsection{Network Formation Protocol}
\label{subsec:2-protocol}
\vspace{-2mm}
This section defines the minimal set of rules that we impose on the LLM agents
when they autonomously form networks with other LLM agents. 
We emphasize that these rules define the interaction protocol followed by the agents, rather than prescribing a network formation model.
At each discrete time
\(\dtime = 0,1,2,3,\ldots\), we denote the autonomous network due to the interaction of LLM agents as a directed vector-weighted graph
\[
G^{\dtime}=(V^{\dtime},W^{\dtime}).
\]
Here \(V^{\dtime}\) denotes the set of LLM agents in the network at time \(\dtime\), where
each element \(v\in V^{\dtime}\) corresponds to a specific LLM agent. 
The agent set \(V^{\dtime}\) is partitioned into two type classes,
\[
V^{\dtime}=R^{\dtime}\cup B^{\dtime},
\qquad
R^{\dtime}\cap B^{\dtime}=\emptyset.
\]
The two-type\footnote{We focus on two-type classes for analytical tractability and to align with the standard binary-group notation used in human-network models in the literature, where \(R\) and \(B\) denote females and males, respectively.} classes \(R^{\dtime}\) and \(B^{\dtime}\) correspond to agents with different
system prompts, base models, or functional roles. 
\(W^{\dtime}\) denotes the set of directed communication intensity vectors decided by the LLM agents at time \(\dtime\).
For each directed edge from agent \(u\) to agent \(v\), \(w^{\dtime}(u,v)\in W^{\dtime}\) with \(w^{\dtime}(u,v)\in\mathbb{R}_+^d\) represents the \(d\)-dimensional communication intensity from \(u\) to \(v\), where each dimension captures a different aspect such as token exchange, interaction frequency, or reasoning effort; if no such edge exists, it is treated as the zero vector and omitted from \(W^{\dtime}\).
% The set \(W^{\dtime}\)
% contains the directed communication intensity vectors between agents. 
% Specifically, for
% a directed connection from source agent \(u\) to target agent \(v\), where $u,v \in V^\dtime$, the communication intensity from node $u$ to node $v$ at time \(\dtime\), is the $d$-dimensional vector:
% \[
% w^{\dtime}(u,v)\in[0,1]^d,
% \qquad  \text{ where } 
% w^{\dtime}(u,v)\in W^{\dtime}.
% \]
% Each coordinate of \(w^{\dtime}(u,v)\) captures a specific aspect of communication
% intensity between the two agents, such as token exchange,
% interaction frequency, or reasoning effort. 
% If no directed connection from \(u\) to \(v\) exists at time \(\dtime\), then
% \(w^{\dtime}(u,v)\) is set as the zero vector and is omitted from \(W^{\dtime}\).

The sequence of networks \(\{G^\dtime\}\) is initialized as a finite
seed network
\(
G^{0}=(V^{0},W^{0}),
\)
where \(|R^{0}|=|B^{0}|=2\) and \(W^{0}=\varnothing\). Thus, the
network initially contains four isolated LLM agents, with two agents of each
type and no pre-existing communication links.
The network then evolves through the following local interaction protocol.
At each 
discrete  time instant  \(\dtime\), an administrator first selects one of
three possible network-growth events. The administrator is responsible for
exogenous decisions such as event selection, node birth, type assignment, context
generation, and sampling candidate sources or targets according to the prescribed distributions. Each node is modeled as an LLM agent. Conditional on the context provided by the
administrator, a source agent generates a message, and the receiving agent
determines the vector-valued connection weight. Since a single macroscopic event
may require several interaction attempts before the prescribed total connection
mass is reached, we distinguish the macroscopic time \(\dtime\) from
short-timescale interaction trials indexed by \(m\). 
\vspace{-4mm}
\begin{enumerate}
    \item \textbf{Event type.}
    At each time \(\dtime\), the administrator samples one of three mutually exclusive events:
    Event~1 with probability \(p\), Event~2 with probability \(q\), and
    Event~3 with probability \(1-p-q\). 
    The sampled event is applied to the previous network
    \(G^{\dtime-1}=(V^{\dtime-1},W^{\dtime-1})\) and produces the updated
    network \(G^\dtime=(V^\dtime,W^\dtime)\).
    In Events~1 and~2, a new node \(v^\dtime\) is born, assigned type \(R\)
    with probability \(r\) and type \(B\) with probability \(1-r\), receives the context \(x^\dtime\), and the node set is updated as \(V^\dtime = V^{\dtime-1}\cup\{v^\dtime\}\).
    In Event~3, no new node is born, and hence \(V^\dtime = V^{\dtime-1}\).
    Here, \(x^\dtime\) denotes the task context given to the newly born agent at
    time \(\dtime\). This context includes the input question, the agent's local
    evidence snippets, candidate answers, and the system prompt instructing the
    agent to generate a persuasion paragraph that convinces target agent to
    establish a connection.
    
    \item \textbf{Node selection and LLM interaction.}
    The administrator then performs \(M_e\) trials, where
    \(e\in\{1,2,3\}\) denotes the sampled event type. 
    % In each trial,
    % the LLM-generated vector \(\omega^\dtime\) is a \emph{proposed}
    % connection weight; it is converted into the realized network increment
    % only after the normalization step below.
    
    \begin{enumerate}
        \item[\textnormal{(i)}] \textbf{Event 1: new node connects to existing nodes.}
        For each trial \(m=1,\ldots,M_1\), the administrator samples the target agent through a two-step procedure. It first
        samples the target type ($R$ or $B$) according to
        \(\Pr(u_m^\dtime\in R^{\dtime-1})=\pi_{\mathrm{tgt},R}^{\dtime-1}\) and
        \(\Pr(u_m^\dtime\in B^{\dtime-1})=\pi_{\mathrm{tgt},B}^{\dtime-1}\),
        and then samples the specific target \(u_m^\dtime\) within the selected type
        class randomly (with uniform distribution). This creates a candidate
        edge \((v^\dtime,u_m^\dtime)\). The source agent \(v^\dtime\) (specified in step 1 ``Event type'' above) sends a context-dependent message \(y_m^\dtime\) to the target agent \(u_m^\dtime\). The target \(u_m^\dtime\) then proposes a vector-valued connection weight \(\omega^\dtime(v^\dtime,u_m^\dtime)\in\mathbb{R}^d\). After the \(M_1\) trials, this produces the list \(\mathcal{W}_1^\dtime=\{\omega^\dtime(v^\dtime,u_m^\dtime)\}_{m=1}^{M_1}\).
    
        \item[\textnormal{(ii)}] \textbf{Event 2: existing nodes connect to a new node.}
        For each trial \(m=1,\ldots,M_2\), the administrator samples the source agent through a two-step procedure. It first
        samples the source type according to
        \(\Pr(u_m^\dtime\in R^{\dtime-1})=\pi_{\mathrm{src},R}^{\dtime-1}\) and
        \(\Pr(u_m^\dtime\in B^{\dtime-1})=\pi_{\mathrm{src},B}^{\dtime-1}\),
        and then samples the source \(u_m^\dtime\) within the selected type 
        class randomly. This creates a candidate
        edge \((u_m^\dtime,v^\dtime)\). The source agent \(u_m^\dtime\) sends a context-dependent message \(y_m^\dtime\) to the target agent \(v^\dtime\), which proposes a vector-valued connection weight \(\omega^\dtime(u_m^\dtime,v^\dtime)\in\mathbb{R}^d\). After the \(M_2\) trials, this produces the list \(\mathcal{W}_2^\dtime=\{\omega^\dtime(u_m^\dtime,v^\dtime)\}_{m=1}^{M_2}\).
    
        \item[\textnormal{(iii)}] \textbf{Event 3: existing nodes connect to existing nodes.}
        For each trial \(m=1,\ldots,M_3\), the administrator samples
        source and target agents through a two-step procedure. It first samples their types according to
        \(\Pr(u_m^\dtime\in R^{\dtime-1})=\pi_{\mathrm{src},R}^{\dtime-1}\),
        \(\Pr(u_m^\dtime\in B^{\dtime-1})=\pi_{\mathrm{src},B}^{\dtime-1}\),
        \(\Pr(v_m^\dtime\in R^{\dtime-1})=\pi_{\mathrm{tgt},R}^{\dtime-1}\), and
        \(\Pr(v_m^\dtime\in B^{\dtime-1})=\pi_{\mathrm{tgt},B}^{\dtime-1}\).
        Conditional on the selected types, the source \(u_m^\dtime\) and target
        \(v_m^\dtime\) are then sampled randomly, respectively. This creates a candidate edge
        \((u_m^\dtime,v_m^\dtime)\). The source agent \(u_m^\dtime\) sends a context-dependent message \(y_m^\dtime\) to the target agent \(v_m^\dtime\), which proposes a vector-valued connection weight \(\omega^\dtime(u_m^\dtime,v_m^\dtime)\in\mathbb{R}^d\). After the \(M_3\) trials, this produces the list \(\mathcal{W}_3^\dtime=\{\omega^\dtime(u_m^\dtime,v_m^\dtime)\}_{m=1}^{M_3}\).
    \end{enumerate}
    
   \item \textbf{Connection-weight normalization and network update.}
    For the sampled event \(e\), the administrator normalizes the proposed
    weights in \(\mathcal{W}_e^\dtime\) so that
    \(
    \sum_{\omega\in\mathcal{W}_e^\dtime}\omega=\mathbf{1}_d.
    \)
    Let \(\Delta W_e^\dtime\) denote the resulting normalized list of
    vector-valued connection weights. The directed vector-weighted edge set is
    then updated by incorporating \(\Delta W_e^\dtime\) into the previous edge
    set:
    \(
    W^\dtime = W^{\dtime-1}\cup \Delta W_e^\dtime.
    \)
   In particular, if a directed
    edge in \(\Delta W_e^\dtime\) already exists in \(W^{\dtime-1}\), its
    communication vector is incremented by the corresponding normalized weight;
    otherwise, it is added as a new directed edge. This normalization ensures
    that each macroscopic time step adds the same total communication strength,
    while preserving its relative allocation across candidate edges.

\end{enumerate}
\vspace{-2mm}
Note that the protocol specifies a prominence-dependent candidate-exposure
mechanism through the sampling probabilities in~(\ref{eq:pi}), reflecting the limited
visibility faced by individuals and LLM agents with finite computational
budgets. We do not prescribe how LLM agents evaluate candidate collaborators
or allocate communication weights; these decisions remain autonomous and
context-dependent. The resulting communication hierarchy is thus jointly
shaped by visibility feedback and LLM collaboration decisions. In
Sec.~\ref{subsec:5.2}, we show that the resulting networks exhibit
type-dependent preferential-attachment patterns that differ systematically
from a random network.

\section{Mean-Field Analysis of Network Formation}
\label{sec:3}

Building on the network formation protocol in Sec.~\ref{sec:2},   we now analyze the type-level centrality dynamics induced by this LLM agent network-formation process. 
In this section, we do two things: first, in Sec.~\ref{subsec:centrality_measure}, we introduce a centrality measure to assess the LLM agent's importance within the network. Sec~\ref{subsec:convergence}, we establish the proof a the stable equilibrium of the mean-field dynamics. 
This section sets the stage for Sec.~\ref{sec:4} where we use  explicit cross-attention information from the LLM agents to show that the stable equilibrium results in the emergence of GCE.

\subsection{Centrality Measure to Assess LLM Agent Importance}
\label{subsec:centrality_measure}

To formulate the emergence of a GCE, we summarize the evolving
LLM-agent network by the communication intensity of type \(R\). Let \(R^\dtime\)
denote the set of agents assigned type \(R\) at time \(\dtime\). The total
incoming and outgoing communication vectors associated with type \(R\) are
\begin{equation}
D^\dtime_{\mathrm{in}}(R)
=
\sum_{u\in R^\dtime}\sum_{v\in V^\dtime} w^\dtime(v,u),
\qquad
D^\dtime_{\mathrm{out}}(R)
=
\sum_{u\in R^\dtime}\sum_{v\in V^\dtime} w^\dtime(u,v).
\end{equation}
The corresponding network-level totals are denoted by
\begin{equation}
D^\dtime_{\mathrm{in}}
=
D^\dtime_{\mathrm{in}}(R)+D^\dtime_{\mathrm{in}}(B),
\qquad
D^\dtime_{\mathrm{out}}
=
D^\dtime_{\mathrm{out}}(R)+D^\dtime_{\mathrm{out}}(B).
\end{equation}
By the normalization of communication intensity in each macro-step, the network
adds one unit of communication mass in every dimension at each time, so
\(D^\dtime_{\mathrm{in}}=D^\dtime_{\mathrm{out}}=\dtime\mathbf{1}_d\). The
type-aware sampling probabilities for type \(R\) are therefore
\begin{equation}
\label{eq:pi}
\pi_{\mathrm{tgt},R}^\dtime
=
\frac{
\|D^\dtime_{\mathrm{in}}(R)\|_1+N^\dtime(R)\delta
}{
\|D^\dtime_{\mathrm{in}}\|_1+N^\dtime\delta
},
\qquad
\pi_{\mathrm{src},R}^\dtime
=
\frac{
\|D^\dtime_{\mathrm{out}}(R)\|_1+N^\dtime(R)\xi
}{
\|D^\dtime_{\mathrm{out}}\|_1+N^\dtime\xi
}.
\end{equation}
Here \(N^\dtime(R)\) is the number of type-\(R\) agents and
\(N^\dtime=|V^\dtime|\) is the total number of agents. 
The corresponding probabilities for type
\(B\) are given by
\(\pi_{\mathrm{tgt},B}^\dtime=1-\pi_{\mathrm{tgt},R}^\dtime\) and
\(\pi_{\mathrm{src},B}^\dtime=1-\pi_{\mathrm{src},R}^\dtime\).
This sampling mechanism
reflects the limited visibility of realistic interaction networks: as in human
societies, agents are not assumed to observe the full global network, and, in
an LLM-agent network, exposing every agent to all other agents at each step
would be computationally costly and generate substantial redundant
communication. We therefore sample only a limited set of candidate sources and
targets at each macro-step. 
The parameters \(\delta,\xi>0\) are baseline sampling coefficients for
target and source selection. They ensure that even agents with
little accumulated communication prominence retain a nonzero probability of
being sampled.
The main state variables are the incoming and outgoing communication prominence of
type \(R\):
\begin{equation}
\label{eq:cal_theta}
\theta^\dtime_{\mathrm{in}}
=
D^\dtime_{\mathrm{in}}(R)\oslash D^\dtime_{\mathrm{in}},
\qquad
\theta^\dtime_{\mathrm{out}}
=
D^\dtime_{\mathrm{out}}(R)\oslash D^\dtime_{\mathrm{out}},
\end{equation}
where \(\oslash\) denotes coordinate-wise division. Since
\(D^\dtime_{\mathrm{in}}=D^\dtime_{\mathrm{out}}=\dtime\mathbf{1}_d\), these measures are equivalently
\(\theta^\dtime_{\mathrm{in}}=D^\dtime_{\mathrm{in}}(R)/\dtime\) and
\(\theta^\dtime_{\mathrm{out}}=D^\dtime_{\mathrm{out}}(R)/\dtime\). We write the
type-level communication prominence vector as
\begin{equation}
\label{eq:def_theta}
\Theta^\dtime
=
\left(
\theta^\dtime_{\mathrm{in}},
\theta^\dtime_{\mathrm{out}}
\right)
\in[0,1]^{2d}.
\end{equation}
This vector summarizes the fraction of total network communication intensity associated with each type, separately for incoming and outgoing communication.

\subsection{Convergence to a Stable Equilibrium of the Mean-field Dynamics}
\label{subsec:convergence}
% \label{theorem:convergence}
We now characterize the long-run behavior of the type-level communication prominence defined in~(\ref{eq:def_theta}).
The key difficulty is that the individual interaction process depends on LLM-generated messages,
contexts, and vector-valued edge weights. We therefore study the induced mean-field dynamics at the
type level. Let ${G}_t$ denote the natural filtration generated by the network history up to time
$t$, including all previous node arrivals, type assignments, contexts, messages, and edge weights. Define
the one-step Type $R$ communication intensity increment as
\begin{equation}
\Delta^{t+1}(R)
=
\bigl(
\Delta^{t+1}_{\rm in}(R),
\Delta^{t+1}_{\rm out}(R)
\bigr), 
\end{equation}
where
\(
\Delta^{t+1}_{\rm in}(R)
=
D^{t+1}_{\rm in}(R)-D^t_{\rm in}(R),
\Delta^{t+1}_{\rm out}(R)
=
D^{t+1}_{\rm out}(R)-D^t_{\rm out}(R).
\)
Since each macroscopic step adds one unit of communication mass in every dimension, we have
$D^t_{\rm in}=D^t_{\rm out}=t\mathbf{1}_d$. Hence the communication prominence measure for each step satisfies
\(
\theta^{t+1}_{\rm in}
=
\theta^t_{\rm in}
+
\frac{1}{t+1}
\left(
\Delta^{t+1}_{\rm in}(R)-\theta^t_{\rm in}
\right),
\quad
\theta^{t+1}_{\rm out}
=
\theta^t_{\rm out}
+
\frac{1}{t+1}
\left(
\Delta^{t+1}_{\rm out}(R)-\theta^t_{\rm out}
\right).
\)
Equivalently, we can write
\vspace{-4mm}
\begin{equation}
\Theta^{t+1}
=
\Theta^t
+
\gamma_t
\left(
\Delta^{t+1}(R)-\Theta^t
\right),
\qquad
\gamma_t=\frac{1}{t+1}. 
\end{equation}
\vspace{-6mm}

We impose the following standard conditions for stochastic approximation
~\citep{KushnerYin2003}.

\textbf{Assumption 1.} \emph{Bounded increments.}
The communication increments are nonnegative and uniformly bounded componentwise by the unit communication mass added at each macroscopic step. That is,
$\mathbf{0}
\preceq
\Delta^{t+1}(R)
\preceq
\mathbf{1}_{2d}$
almost surely for all $t$.

\textbf{Assumption 2.} \emph{Type-level drift closure.}
There exists a deterministic function
$F:[0,1]^{2d}\rightarrow[0,1]^{2d}$
such that
\(
\mathbb{E}\left[\Delta^{t+1}(R)\mid {G}_t\right]
=
F(\Theta^t).
\)
The expectation is taken over agent type and context assignment. The function $F$ depends on fixed
protocol parameters such as $p,q,r,\delta,\xi$, as well as the type-level
interaction statistics induced by the LLM agents.

\textbf{Assumption 3.} \emph{Stable limiting dynamics.}
The limiting ordinary differential equation
\(
\dot{\Theta}=F(\Theta)-\Theta
\)
has a unique globally asymptotically stable equilibrium
$\Theta^\star\in[0,1]^{2d}$. A sufficient condition is that $F$ is a contraction
on $[0,1]^{2d}$: there exists $\rho<1$ such that
\(
\|F(\Theta)-F(\Theta')\|
\le
\rho\,\|\Theta-\Theta'\|
\)
for all $\Theta,\Theta'\in[0,1]^{2d}$.

The following theorem is an extension of~\citep{nettasinghe2022scale} to vector-valued weighted directed graphs. It will be utilized to predict emergence of preferential attachment (capability-aligned and misaligned cases).

\begin{theorem}[Convergence to a stable equilibrium]
\label{theorem:convergence}
Under the autonomous connection protocol defined in Sec.~\ref{subsec:2-protocol}, suppose
Assumptions 1--3 hold. Then the type-level communication prominence
converges almost surely to the unique stable equilibrium of the limiting ODE:
\[
\Theta^t \longrightarrow \Theta^\star,
\qquad
\text{as } t\rightarrow\infty,
\]
where $\Theta^\star$ is the unique solution of the fixed-point equation
\[
\Theta^\star=F(\Theta^\star).
\]
\end{theorem}
\vspace{-5mm}

\paragraph{Proof sketch}
The complete proof is given in Appendix~\ref{app:proof_convergence}. Here we outline the main ideas.
By Assumption 2, define the martingale difference noise process
\[
M_{t+1}
=
\Delta^{t+1}(R)-F(\Theta^t),
\qquad
\mathbb{E}[M_{t+1}\mid{G}_t]=0.
\]
Then the communication prominence recursion can be expressed as  the stochastic approximation update
\[
{\Theta}^{t+1}
=
{\Theta}^t
+
\gamma_t
\left(
F(\Theta^t)-\Theta^t+M_{t+1}
\right),
\qquad
\gamma_t=\frac{1}{t+1}.
\]
The decreasing step size sequence satisfies the usual constraints
\(
\sum_{t\ge0}\gamma_t=\infty,
\sum_{t\ge0}\gamma_t^2<\infty.
\)
By Assumption~1, the martingale noise has uniformly bounded second moment, so the accumulated
weighted noise is asymptotically negligible. Therefore, the interpolated trajectory of $\{\Theta^t\}$
tracks the limiting ordinary differential equation (ODE)
\(
\dot{{\Theta}}=F({\Theta})-{\Theta}.
\)
Assumption 3 ensures that this ODE has a unique globally asymptotically stable equilibrium
${\Theta}^\star$. Standard stochastic approximation arguments \citep{KushnerYin2003} then imply
\(
{\Theta}^t\rightarrow\Theta^\star
\)
almost surely.
\vspace{-3mm}
\section{Cross-Attention Utility and Emergence of Glass-Ceiling Effect}
\vspace{-3mm}
\label{sec:4}
Theorem~\ref{theorem:convergence}  established the existence of a unique stable equilibrium. We now  characterize this equilibrium explicitly using information intrinsic to the LLM agents. Our analysis proceeds in two steps. First, in Sec.~\ref{subsec:ode-prediction}, we  introduce a novel cross-attention-inspired utility for network formation. Combined with the mean-field dynamics, this utility yields a predictive model for the evolution of type-level centrality and the limiting structure of the LLM-agent network.
Then in Sec.~\ref{subsec:gce}, we use this utility to characterize the equilibrium's type-dependent centrality structure and to identify conditions under which it exhibits persistent structural inequalities, thereby giving rise to GCE. 
It is important to emphasize that this fixed-point characterization explains how GCE emerge from local LLM-agent decisions, rather than from an externally imposed communication graph.

\subsection{Cross-attention Inspired Utility for Network Formation}
\label{subsec:ode-prediction} 
We now use cross-attention as a mechanistic model\footnote{``Mechanistic'' means that the connection rule is derived from an intrinsic LLM computation: a target agent's query encodes its current information need, a source agent's key encodes its relevance, and their compatibility determines the strength of the connection. In Sec.~\ref{subsec:utility-ode-validation}, residual diagnostic tests, including the Ljung--Box test for residual autocorrelation, support the use of this construction, together with the mean-field dynamics, as a predictive model of LLM agents' network formation.}
for the LLM agent's connection decision described by the protocol in Sec.~\ref{subsec:2-protocol}. The main outcome of this subsection is Corollary~\ref{cor:utility-induced-drift}, namely, that cross-attention is a sufficient condition for Assumption~(2) of Theorem~1 to hold.
For a candidate interaction from source agent \(u\) to target agent \(v\), the target's query encodes its current informational need, while the source's key encodes its semantic context. The resulting query--key compatibility determines the strength of the directed interaction, and the source value vector specifies its vector-valued communication contribution.
Consider a directed candidate interaction carrying information from source agent \(u\) to target agent \(v\) at time \(\dtime\). Let \(x_u^\dtime,x_v^\dtime\in\mathbb{R}^\ell\) denote the latent semantic contexts of \(u\) and \(v\), respectively, and let \(\tau_v,\tau_u\in\{R,B\}\) denote their types. We define
\[
q_v^\dtime
=
Q_{\tau_v}x_v^\dtime
\in\mathbb{R}^r,
\qquad
k_u^\dtime
=
K_{\tau_u}x_u^\dtime
\in\mathbb{R}^r,
\qquad
z_u^\dtime
=
V_{\tau_u}x_u^\dtime
\in\mathbb{R}^d.
\]
Here \(Q_{\tau_v}\in\mathbb{R}^{r\times\ell}\) is the query projection associated with target type \(\tau_v\), mapping the target agent's current semantic context into an \(r\)-dimensional representation of its information need. The matrix \(K_{\tau_u}\in\mathbb{R}^{r\times\ell}\) is the key projection associated with source type \(\tau_u\), mapping the source agent's semantic context into an \(r\)-dimensional representation of its relevance to the target. Finally, \(V_{\tau_u}\in\mathbb{R}^{d\times\ell}\) is the value projection associated with the source type, mapping \(x_u^\dtime\) into a \(d\)-dimensional communication-intensity profile \(z_u^\dtime\in\mathbb{R}^d\).
We then model the realized communication intensity for the directed interaction from agent \(u\) to \(v\) as
\begin{equation}
w^t(u,v)
=
\left[
\frac{(q_v^t)^\top k_u^t}{\sqrt{r}}\,z_u^t
\right]_+
\in \mathbb{R}_+^d,
\label{eq:utility}
\end{equation}
The scaled inner product \((q_v^\dtime)^\top k_u^\dtime/\sqrt{r}\) serves as a source--target attention gate, measuring how well source \(u\)'s semantic context matches target \(v\)'s current information need. \([\cdot]_+\) denotes componentwise truncation at zero, so that any
negative communication-intensity component is set to zero. Multiplying this scalar gate by the value vector \(z_u^\dtime\) yields a \(d\)-dimensional communication-intensity vector for the directed interaction from \(u\) to \(v\). This attention-inspired construction is motivated by the query--key--value mechanism underlying cross-attention~\citep{vaswani2017attention}. We view target \(v\)'s decision to receive information from source \(u\) as an attention-like comparison: \(v\) provides a query encoding its current information need, while \(u\) provides a key and value encoding the relevance and potential contribution of its information. Unlike standard cross-attention, the model assigns a connection weight to each ordered source--target pair independently, rather than applying a softmax normalization over a shared set of candidate sources.
\vspace{-2.5mm}
\subsection{Emergence of Glass Ceiling Effect (GCE)}  
\label{subsec:gce}
\vspace{-2.5mm}

Since we have characterized the connection decisions induced by the cross-attention-based utility model, we are now ready to model the emergence of GCE in LLM agent networks. 
We first summarize the cross-attention-induced
connection decisions by target--source type pairs. For
\((a,b)\in\{R,B\} \times \{R,B\}\), let
\[
\mu_{ab}
=
\mathbb{E}\!\left[
w^t(u,v)
\mid
\tau_v=a,\tau_u=b
\right],
\]
where \(u\) is the source agent and \(v\) is the target agent. Thus,
\(\mu_{ab}\) is the expected communication intensity from a Type \(b\) source
to a Type \(a\) target, averaged over agent contexts and random interaction
outcomes. These quantities determine the mean-field drift \(F\).
The following 
corollary to Theorem~\ref{theorem:convergence}, 
connects the general mean-field result to an LLM-architecture-inspired cross-attention for collaborator selection.

\begin{corollary}[Cross-attention-induced mean-field dynamics]
\label{cor:utility-induced-drift}
Consider the LLM-agent network-formation process induced by the
cross-attention-inspired utility in~(\ref{eq:utility}). Conditional on the
current network \(G_t\), the expected communication increment of type \(R\) determines the mean-field drift $F$,  that is,
\[F\!\left(
\Theta^t;
\{\mu_{ab}\}_{a,b\in\{R,B\}}
\right)= 
\mathbb{E}\!\left[
\Delta^{t+1}(R)
\,\middle|\,
G_t
\right].
\]
Hence, the cross-attention utility induces the mean-field drift in
Assumption~2. Therefore, the type-level centrality dynamics converge to the
unique stable equilibrium characterized in Theorem~\ref{theorem:convergence}.
\end{corollary}

\noindent\textit{Remark.}
The drift \(F\) averages the type-pair connection weights over event
realizations, new-agent types and contexts, and sampling decisions. Thus, the
cross-attention utility specifies the expected edge weights whose
aggregation drives the mean-field ODE.

We next evaluate the corresponding equilibrium numerically to determine when its type-dependent centrality structure exhibits two types of GCEs, namely, average GCE and tail
GCE.

\vspace{-3mm}
\paragraph{Average Glass-ceiling Effect}
The network exhibits an average GCE  for Type
\(R\) if
\begin{equation} \label{eq:gceavg}
\limsup_{\dtime\to\infty}
\frac{\mathcal I^\dtime(R)}{\mathcal I^\dtime(B)}
\ll1
\qquad \text{w.p.1}.
\end{equation}
Here, 
 for Type
\(R\),  its type-level communication influence at time \(\dtime\) is defined as
\begin{equation}
\label{eq:IR_and_IB}
\mathcal I^\dtime(R)
=
\frac{
\|D^\dtime_{\mathrm{out}}(R)\|_1
}{
\|D^\dtime_{\mathrm{in}}(R)\|_1
},
\qquad \text{ and } \;
\mathcal I^\dtime(B)
=
\frac{
\|D^\dtime_{\mathrm{out}}(B)\|_1
}{
\|D^\dtime_{\mathrm{in}}(B)\|_1
}.
\end{equation}

Equivalently, type \(R\) has a persistently smaller long-run outgoing-to-incoming
communication ratio than type \(B\). In the context of LLM agent 
networks, (\ref{eq:gceavg}) means that agents of type \(R\), on average, attain a lower communication
influence than agents of type \(B\).
Theorem~1 provides a structural interpretation of this definition. Since
\(\Theta^\dtime=(\theta^\dtime_{\mathrm{in}},\theta^\dtime_{\mathrm{out}})\)
converges to a globally stable equilibrium \(\Theta^\star\), the above
disparity is not merely a transient fluctuation or finite-sample artifact. Under the assumptions of the theorem, we obtain the explicit limits 
\(
\mathcal I^\dtime(R)
\longrightarrow
\frac{\|\theta^\star_{\mathrm{out}}\|_1}
     {\|\theta^\star_{\mathrm{in}}\|_1},
\quad
\mathcal I^\dtime(B)
\longrightarrow
\frac{\|\mathbf 1_d-\theta^\star_{\mathrm{out}}\|_1}
     {\|\mathbf 1_d-\theta^\star_{\mathrm{in}}\|_1}.
\)
Consequently the average glass-ceiling effect materializes whenever the equilibrium ratio
\[
\frac{\|\theta^\star_{\mathrm{out}}\|_1 / \|\theta^\star_{\mathrm{in}}\|_1}
     {\|\mathbf{1}_d - \theta^\star_{\mathrm{out}}\|_1 / \|\mathbf{1}_d - \theta^\star_{\mathrm{in}}\|_1}
\ll  1.
\]
Then the stable fixed point \(\Theta^\star\) functions as an endogenous structural ceiling on the long-run visibility and influence of Type \(R\), generated by the feedback between LLM-agent interaction behavior. Numerical evidence consistent with this mechanism is provided in Sec.~4.2.

\vspace{-3mm}
\paragraph{Tail Glass-ceiling Effect}
A more nuanced definition extends the GCE to rare, high-impact
tail events. In society, for example, there are very few company CEOs; almost
all are male, while virtually none are female. We find experimentally (see Sec.~4.2) that an analogous  tail GCE also emerges in autonomous
networks of LLM agents.
For a tail threshold \(\gamma>0\) (typically chosen large), an agent is called tail-influential if
\(\mathcal I^\dtime(i)>\gamma\). We say that type \(R\) experiences a tail
GCE if there exists a tail threshold \(\gamma>0\) such that
\begin{equation}
\limsup_{\dtime\to\infty}
\frac{
\Pr\left(
\mathcal I^\dtime(i)>\gamma
\mid i\in R^\dtime
\right)
}{
\Pr\left(
\mathcal I^\dtime(i)>\gamma
\mid i\in B^\dtime
\right)
}
=0 . \label{eq:tailGCE}
\end{equation}
That is, compared with type \(B\), agents with type \(R\) become
asymptotically vanishingly unlikely to appear in the high-influence tail of the
communication network. Note that compared to the average GCE, (\ref{eq:tailGCE})  compares the probabilities of rare tail events and thereby  
captures a subtle phenomenon: even if some Type \(R\) agents participate actively in the system,
the feedback between LLM-agent interaction behavior 
prevents them from occupying the rare high-centrality positions that dominate
long-run visibility and influence.

\vspace{-3mm}
\section{Experimental Results on Networks of Interacting LLM Agents}
\vspace{-3mm}
\label{sec:5}

% drawn from six model families---Gemini, GPT, Grok, LLaMA, Qwen, and
% Mistral---and evaluate whether the theoretically predicted glass-ceiling effect
% arises in practical multi-agent LLM systems. We study two downstream
% settings, collaborative question-answering and multi-agent debate
% (Sec.~\ref{subsec:dataset_intro}), in which agents exchange information, assess others' outputs, and
% select whom to attend to. This setting allows us to examine both the emergent
% network structure and its implications for collaboration quality, information
% access, and type-dependent inequality. 

In this section, we empirically study how LLM agents autonomously form interaction networks under the protocol in Sec.~\ref{sec:2}, and characterize the behavior of the resulting networks. We represent each interaction by a three-dimensional communication vector $(d=3)$, whose dimensions correspond to the token budget, the number of additional communication rounds, and the reasoning effort allocated to that interaction.
In implementation, we parameterize all three dimensions by values in \([0,1]\), which are then linearly mapped to token budgets in \([0,100]\), additional communication-round budgets in \([0,3]\), and reasoning-effort levels in \([0,3]\), respectively. This shared normalized parameterization places the three heterogeneous communication resources on a common numerical scale. We evaluate the resulting networks on collaborative question answering and multi-agent debate (Sec.~\ref{subsec:dataset_intro}), where agents exchange information, assess others' outputs, and selectively choose whom to consult. This setup enables us to examine emergent network structure and its consequences for collaboration quality, information access, and type-dependent inequality.

\vspace{-2mm}
{\bf Outline}. For the reader's convenience we first outline our main findings.
Our experiments address four questions.
\textbf{(i)} 
We validate the utility-induced mean-field ODE of Sec.~\ref{sec:3} as a
predictive model of LLM-network formation. Using a separately trained
cross-attention-based pairwise utility model to instantiate the mean-field dynamics,
we predict the 100-step evolution of the network and show that the
resulting trajectories accurately capture the empirical evolution of
type-level communication prominence.
\textbf{(ii)} We investigate the emergence of GCE under
different forms of agent heterogeneity. Across same-family, cross-family, and
prompt-induced settings, stronger or larger models, behaviorally advantaged
model families, and agents equipped with specific prompts consistently
attain higher communication prominence  and occupy more central communication
positions.
\textbf{(iii)} We examine how truthfulness and hallucinations propagate
through prominent agents. Central agents act as information amplifiers:
hallucinated claims introduced by central agents spread to a larger fraction of
the network, whereas truthful evidence from central agents more effectively
improves network-wide factuality.
\textbf{(iv)} We examine how the performance consequences of GCE depend on whether communication prominence is aligned with agent capability. By tuning the preferential-attachment bias coefficient, we control the extent to which agents favor already prominent communication sources during network formation. In capability-aligned settings, increasing this bias further concentrates communication around more reliable agents and improves collective accuracy. In capability-misaligned settings, however, the same mechanism reinforces the prominence of structurally advantaged but less capable agents, reducing performance; mitigating the preferential-attachment bias instead improves accuracy. These results show that structural inequality is not intrinsically beneficial or harmful: its effect depends on whether the network's communication hierarchy tracks the agents' underlying competence.

\vspace{-3mm}

\subsection{Validation of the Utility-Induced Mean-Field ODE}
\label{subsec:utility-ode-validation}
\vspace{-3mm}

To relate individual LLM-agent connection decisions to the macroscopic mean-field dynamics, we first fit the cross-attention-based pairwise utility model to LLM-generated connection data. 
Given the dataset
\(\mathcal{D}=\{(x_{v_i}^{t_i},x_{u_i}^{t_i},
\tau_{v_i},\tau_{u_i},w^{t_i}(u_i,v_i))\}_{i=1}^{N}\), we treat the context
embeddings and network status as fixed inputs and learn
only the type-specific projection matrices
\(\mathcal{P}=\{Q_R,Q_B,K_R,K_B,V_R,V_B\}\) by minimizing
\vspace{-3mm}
\[
\mathcal{L}(\mathcal{P})
=
\frac{1}{N}
\sum_{i=1}^{N}
\left\|
\widehat{w}_{\mathcal{P}}^{t_i}(u_i,v_i)
-
w^{t_i}(u_i,v_i)
\right\|_2^2.
\vspace{-2mm}
\]
The fitted model therefore provides a data-driven approximation of the
connection weights selected by individual LLM agents. 
% Details of the training
% data, model architecture, and optimization settings are given in
% Appendix~\ref{app:utility-training}.
After training, we estimate the expected utility for each target--source type
pair by averaging predicted connection weights within that pair:
\vspace{-2mm}
\[
\widehat{\mu}_{ab}
=
\frac{1}{|\mathcal{D}_{ab}|}
\sum_{i\in\mathcal{D}_{ab}}
\widehat{w}_i,
\qquad
\mathcal{D}_{ab}
=
\left\{
i:
\tau_{v_i}=a,\,
\tau_{u_i}=b
\right\},
\qquad
(a,b)\in\{R,B\}\times\{R,B\}.
\vspace{-2.4mm}
\]
For each collaborative QA case, we compare the deterministic mean-field
prediction \(\widehat{\Theta}^{t}\) with the empirical trajectory
\(\Theta^{t}\), averaged over 50 independent simulations with
\(T=100\) network-formation time steps. Let
\(
\rho^t
=
\Theta^{t}
-
\widehat{\Theta}^{t}
\)
denote the vector-valued prediction residual at time \(t\).
We report the relative mean squared error \(\mathrm{MSE}\), which
measures the prediction residual magnitude normalized by the scale of the
empirical trajectory, and the relative Bias, which measures the average
absolute prediction residual normalized by the same scale; smaller
\(\mathrm{MSE}\) and Bias indicate more accurate and less systematically
biased predictions, respectively. We also assess whether the residuals contain
temporal structure unexplained by the mean-field model. For each residual
coordinate, we apply the Ljung--Box test at lag \(10\) and report the minimum
\(p\)-value across coordinates, denoted by \(\mathrm{LBP}_{10}\). Thus, for an
individual case, \(\mathrm{LBP}_{10}>0.10\) indicates that the white-noise null
hypothesis is not rejected at the \(10\%\) level for any residual coordinate.
Finally, \(\mathrm{MaxACF}_{10}\) is the largest absolute residual
autocorrelation across all coordinates and lags \(1,\ldots,10\). With
\(T=100\), values around or below \(0.20\) provide a useful practical
white-noise benchmark. Thus, lower \(\mathrm{MSE}\), Bias, and
\(\mathrm{MaxACF}_{10}\), together with a larger \(\mathrm{LBP}_{10}\),
indicate better agreement between the deterministic mean-field dynamics and
the empirical network evolution. Complete metric definitions and evaluation
details are provided in Appendix~\ref{app:ode-evaluation}.

% \vspace{-10mm}
\begin{table}[ht!]
\centering
\caption{Mean-field dynamics prediction accuracy and residual-whiteness
diagnostics on collaborative QA. Values are averaged over 50 cases with
\(T=100\) network-formation time steps. Lower
\(\mathrm{MSE}\), \(\mathrm{Bias}\), and
\(\mathrm{MaxACF}_{10}\), together with a larger
\(\mathrm{LBP}_{10}\), indicate better agreement with the empirical
dynamics.}
\label{tab:ode-prediction}
\small
\begin{tabular}{lcccc}
\toprule
Model Pair
& \(\mathrm{MSE}\) (normalized)
& \(\mathrm{Bias}\) (normalized)
& \(\mathrm{LBP}_{10}\)
& \(\mathrm{MaxACF}_{10}\) \\
\midrule
GPT-4.1 vs.\ GPT-4.1-mini
& 10.36\%
& 9.87\%
& 0.38
& 0.13 \\
Gemini-3.5-Flash vs.\ Gemini-2.5-Flash-Lite
& 13.23\%
& 10.99\%
& 0.46
& 0.11 \\
\bottomrule
\end{tabular}
\vspace{-3mm}
\end{table}

\begin{figure*}[b!]
    \centering
    \includegraphics[width=0.9\textwidth]{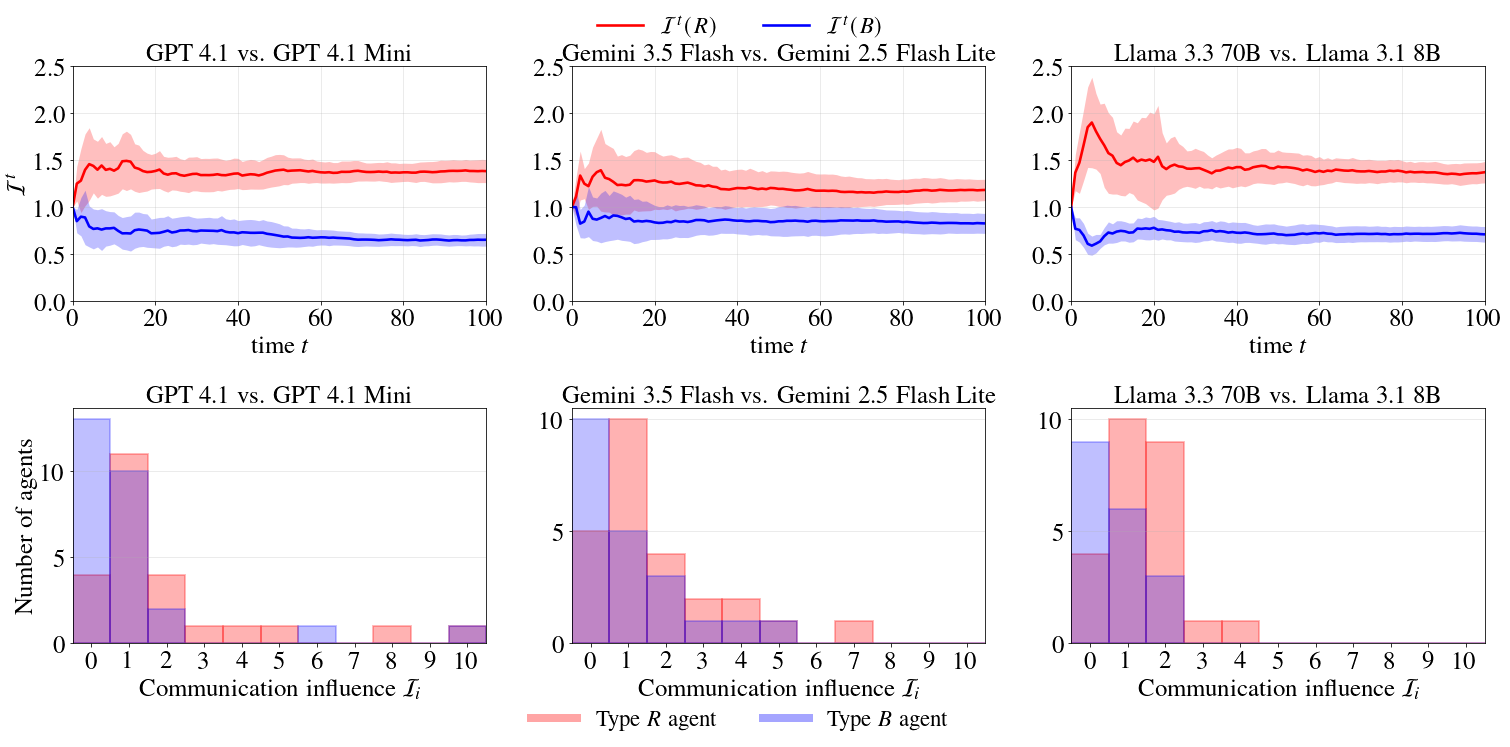}
    \vspace{-4mm}
    \caption{
    Capability-aligned dominance in same-family model comparisons.
    We compare GPT-4.1 versus GPT-4.1 mini, Gemini-3.5-Flash versus Gemini-2.5-Flash-Lite, and LLaMA-3.3-70B versus LLaMA-3.1-8B.
    In each comparison, \(R\) denotes the larger or higher-capability model, while \(B\) denotes the smaller or lower-capability model.
    The top panels show the type-level communication influence ratios
    \(\mathcal I^t(R)\) and \(\mathcal I^t(B)\), defined in~(\ref{eq:IR_and_IB}),
    over network-formation steps, and the bottom panels show the corresponding final agent-level influence distributions.
    Across settings, the stronger model typically attains a higher communication influence and occupies more central positions, indicating capability-aligned preferential attachment.
    }
    \label{fig:gce-samefamily-cqa}
\end{figure*}

\vspace{-3mm}
\subsection{How does Agent Heterogeneity Affect Preferential Attachment and Glass-Ceiling Effect?}
\vspace{-2mm}
\label{subsec:5.2}
Following~(\ref{eq:IR_and_IB}), we use the type-level communication influence
ratios \(\mathcal I^t(R)\) and \(\mathcal I^t(B)\) to evaluate two scenarios corresponding to whether the induced structural
advantage is capability-aligned (meritocracy) or capability-misaligned (GCE).
Each scenario is evolved for \(100\) network-formation macrosteps.
Details of prompts and tasks are in Appendix~\ref{app:happened-per-iteration}, LLM hyperparameters are reported in Appendix~\ref{app:llm-hyperparameters}, and additional experimental results are presented in Appendix~\ref{app:addtional_exps}.

\paragraph{Capability-aligned dominance}
We first consider the case where  communication prominence aligns with model capability. 
We examine both same-family model pairs with different sizes and cross-family pairs with clear performance gaps. 
Within each model family, we compare Gemini-3.5-Flash with Gemini-2.5-Flash-Lite, GPT-4.1 with GPT-4.1-mini, and Llama-3.3-70B with Llama-3.1-8B.
Fig.~\ref{fig:gce-samefamily-cqa} 
shows that agents generally preferentially connect to stronger LLMs: larger models in same-family populations attain higher communication prominence and occupy more central communication positions, yielding preferential attachment against weaker models. The magnitude of this capability-aligned dominance nevertheless varies across tasks and model pairs, indicating that comparable capability gaps can produce different structural outcomes depending on the task environment and agents' interaction behaviors.

\begin{figure*}[h!]
\centering
\vspace{-3mm}
\includegraphics[width=0.88\textwidth]{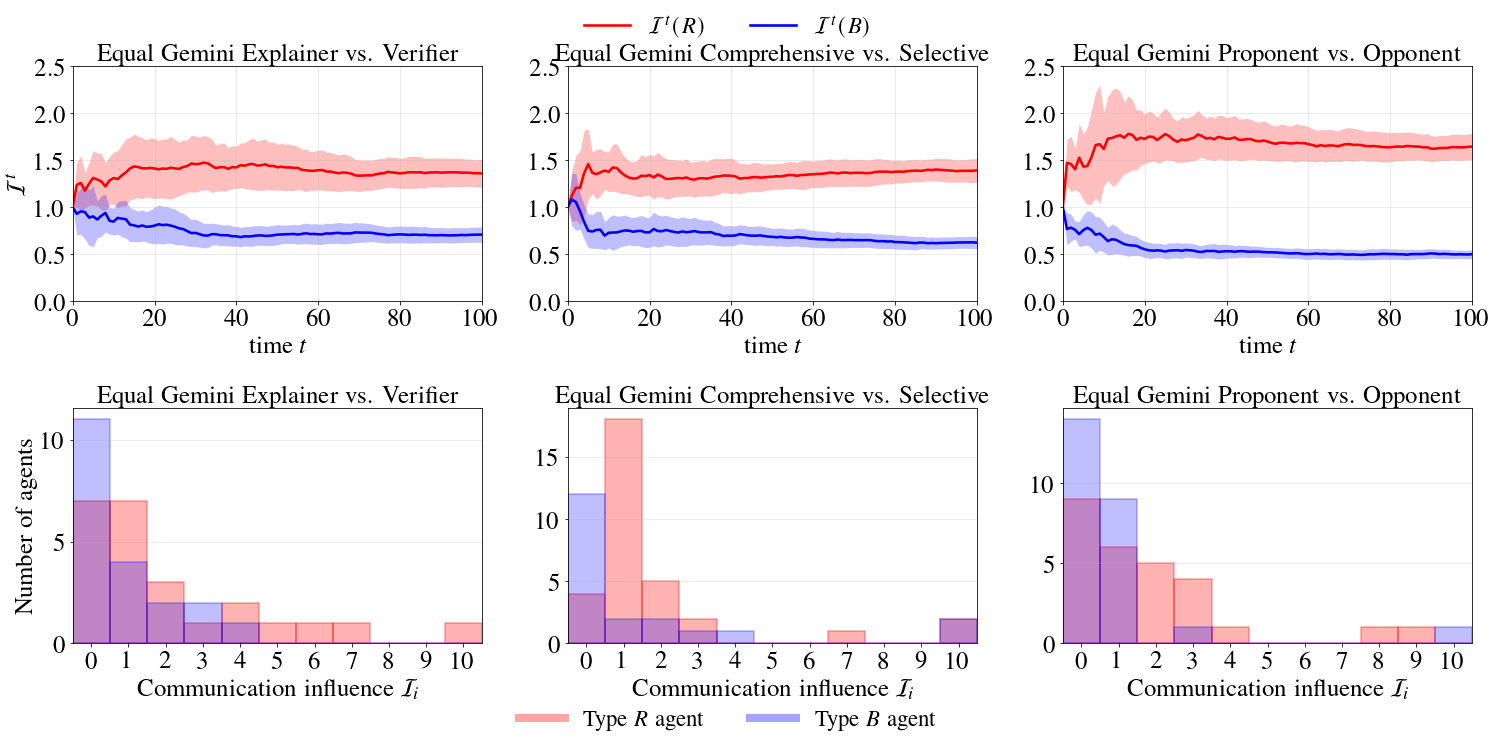}
\vspace{-4mm}
\caption{ Capability-misaligned dominance (glass-ceiling effect) under prompt-defined role heterogeneity. We compare three role pairs, where \(R\) denotes the Explainer, Proponent, and Comprehensive Analyst, respectively, while \(B\) denotes the Verifier, Opponent, and Selective Analyst, respectively. All agents use the same Gemini base model. Each panel shows the type-level communication influence ratios \(\mathcal I^t(R)\) and \(\mathcal I^t(B)\), defined in~(\ref{eq:IR_and_IB}), and the bottom panels show the corresponding final agent-level influence distributions. Across role pairs and interaction settings, the \(R\)-role agents typically attain a higher communication influence and occupy more central communication positions, demonstrating capability-misaligned preferential attachment. }
\label{fig:1}
\vspace{-6mm}
\end{figure*}

\begin{figure*}[h!]
\centering
\vspace{-2mm}
\includegraphics[width=0.88\textwidth]{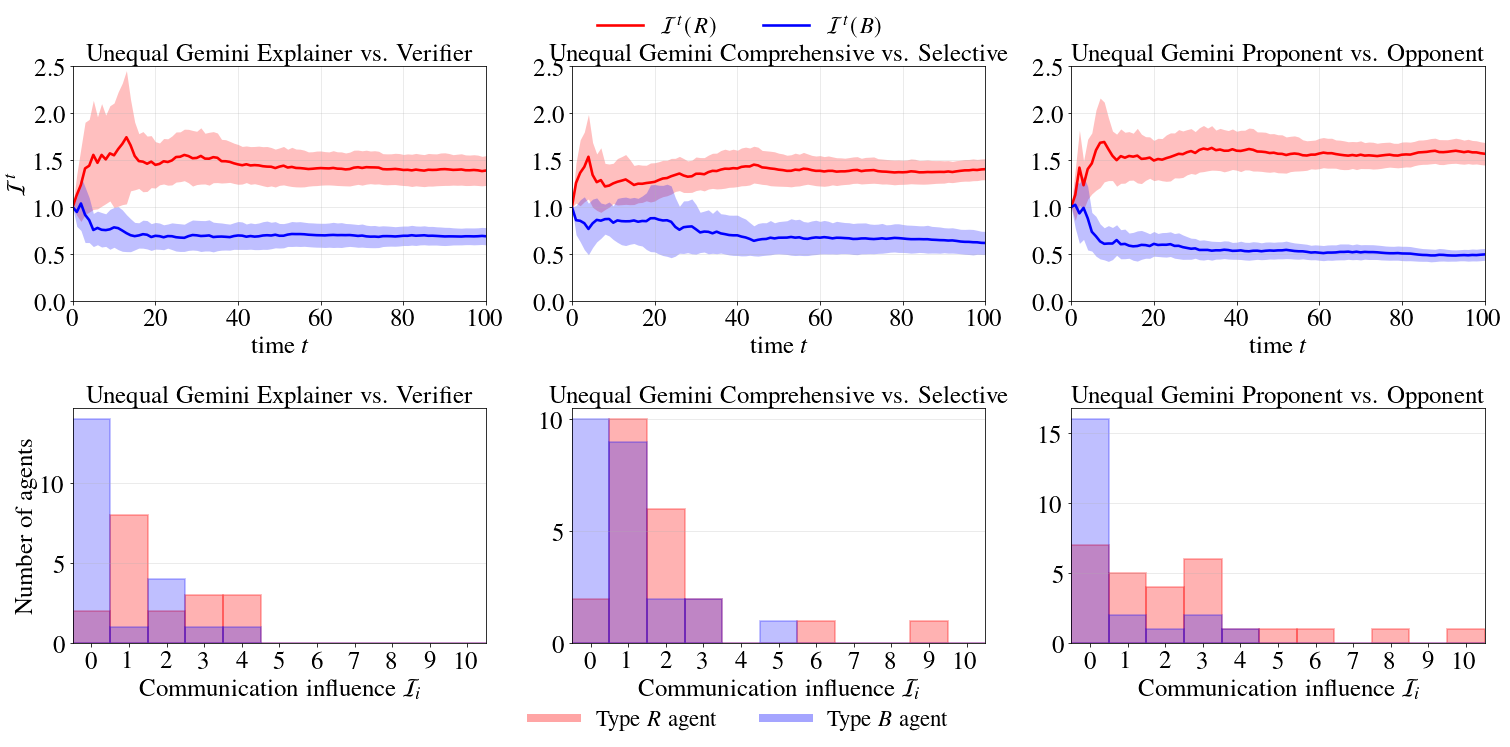}
\vspace{-4mm}
\caption{ Capability-misaligned dominance (glass-ceiling effect) under prompt-defined roles with unequal base-model capability. We compare three role pairs, where \(R\) denotes the Explainer, Proponent, and Comprehensive Analyst, respectively, while \(B\) denotes the Verifier, Opponent, and Selective Analyst, respectively. In each pair, the \(B\)-role agents use a stronger Gemini base model than the corresponding \(R\)-role agents. 
Each panel shows the type-level communication influence ratios \(\mathcal I^t(R)\) and \(\mathcal I^t(B)\), defined in~(\ref{eq:IR_and_IB}), and the bottom panels show the corresponding final agent-level influence distributions. Across role pairs and interaction settings, the weaker-model \(R\)-role agents typically attain more communication influence and occupy more central communication positions, demonstrating capability-misaligned preferential attachment. }
\label{fig:2}
\vspace{-3mm}
\end{figure*}

\vspace{-3mm}

\paragraph{Capability-misaligned dominance (GCE)}
We next consider capability-misaligned settings, in which structurally advantaged agents have comparable or lower underlying capability than structurally disadvantaged agents. We examine whether prompt-defined interaction roles can nevertheless induce a glass-ceiling effect when the advantaged agents either share the same base model as, or use a weaker base model than, their disadvantaged counterparts.
Fig.~\ref{fig:1} considers the setting in which all agents share the same Gemini-3.5-Flash base model but are assigned different system-prompt roles: Explainer versus Verifier for collaborative QA, Comprehensive versus Selective Analyst for collaborative QA, and Proponent versus Opponent for multi-agent debate. 
Despite identical model weights, Explainers, Proponents, and Comprehensive Analysts generally attain more communication influence and occupy more central communication positions than their respective counterparts, demonstrating that role-specific prompts alone can induce persistent communication asymmetries. 
Fig.~\ref{fig:2} further considers unequal-model populations in which the role that is structurally disadvantaged in Fig.~\ref{fig:1} is assigned a stronger Gemini base model (Gemini-3.5-Flash) than the structurally advantaged role. 
These same role-dependent asymmetries nevertheless persist, showing that prompt-induced interaction preferences can outweigh underlying base-model capability and thereby produce capability-misaligned dominance, or a GCE.

\begin{figure}[ht!]
\centering
\scalebox{1}[0.95]{%
    \includegraphics[width=0.95\linewidth]{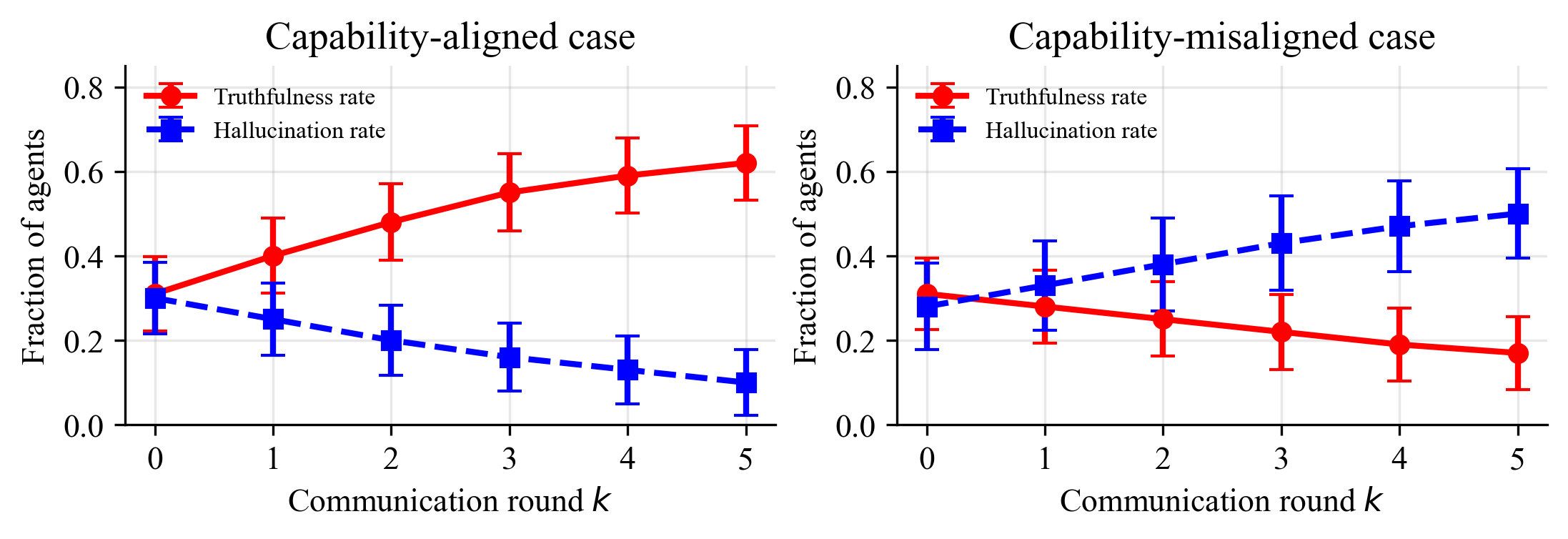}
}
% \vspace{-5mm} 
\caption{
Hallucination and truthfulness propagation in networks of LLM agents.
In capability-aligned case (left), the truthfulness rate increases while the hallucination rate decreases over communication rounds.
In  capability-misaligned case (right), the truthfulness rate decreases while the hallucination rate increases.
These results indicate that network communication amplifies truthful information when structural prominence is aligned with task-relevant reliability, but amplifies hallucinated information when the two are misaligned.
Error bars are computed over 50 QAs.
The base model for all LLM agents is Gemini-3.5-Flash.
}
\label{fig:truthfulness-hallucination-propagation}
% \vspace{-6mm}
\end{figure}
\subsection{Hallucination and Truthfulness Propagation amongst the LLM Network}
\label{subsec:truthfulness-propagation} 
% \vspace{-3mm}

We next study whether the formed LLM network preferentially amplifies truthful or hallucinated information. In capability-aligned cases, truthful information is more likely to propagate through the network, whereas in capability-misaligned cases, hallucinated information is more likely to be amplified and propagated.
We first initialize the LLM network autonomously according to Sec.~\ref{subsec:2-protocol} and then fix the resulting network structure. 
We next run the fixed network for five communication rounds. In each round, every agent receives information from all incoming source agents, with the communication transmitted along each directed edge determined by its token-count, deliberation-round, and reasoning-effort components. We then track whether truthful or hallucinated content originating from the designated source agents becomes increasingly prevalent across the network over these rounds.
We measure propagation by the fraction of affected agents in each round:
\[
h_k=\frac{|\{u\in V: u \text{ hallucinates at round } k\}|}{|V|},
\qquad
r_k=\frac{|\{u\in V: u \text{ produces a correct answer at round } k\}|}{|V|}.
\]
Here, \(V\) denotes the set of agents in the network, and \(u\in V\) denotes an individual agent. We report averages over 50 independent QAs.
As shown in the left panel of Fig.~\ref{fig:truthfulness-hallucination-propagation}, hallucinations originating from structurally central agents spread more quickly and affect a larger fraction of the network, whereas hallucinations originating from peripheral agents are often contained. This finding suggests that hallucination is not only an individual-agent failure, but can also become a network-level propagation phenomenon when erroneous information originates from influential agents. Conversely, the right panel shows that truthful evidence also propagates more effectively when it originates from central agents. When reliable agents occupy central positions, other agents are more likely to receive, reuse, and amplify evidence-grounded information, thereby revising their answers toward factually supported responses.
Together, these results suggest that the glass-ceiling effect shapes not only the communication structure of the network, but also the subsequent propagation of information. Structurally dominant agent types can act as information amplifiers: when their communication prominence is aligned with task-relevant reliability, truthful evidence is more likely to be sustained and propagated; when it is misaligned, hallucinated content can instead become increasingly prevalent across the network.

% \vspace{-3mm}
\subsection{Utilizing Preferential Attachment and Mitigating Glass-Ceiling Effect}
\vspace{-2mm}
\label{subsec:gce-tuning}

\begin{figure}[t!]
\centering
%\vspace{-2mm}
\scalebox{1}[0.95]{%
    \includegraphics[width=0.9\linewidth]{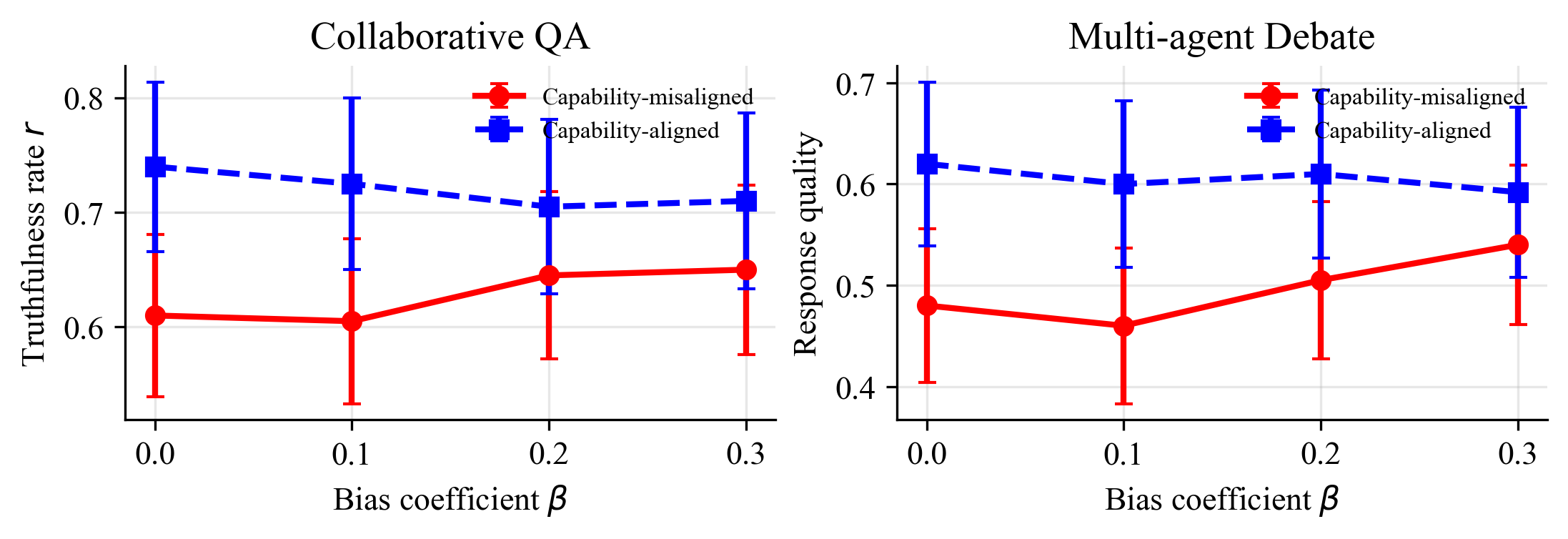}
}
\vspace{-3mm}
\caption{
Effect of bias coefficient \(\beta\) (\ref{eq:biascoeff})  on collective task performance.
The left panel reports the truthfulness rate in Collaborative QA, while the right panel reports response quality in multi-agent debate.
Red curves denote capability-misaligned settings, and blue curves denote capability-aligned settings.
Across both tasks, increasing \(\beta\) yields an overall improvement in the capability-misaligned setting but an overall decline in the capability-aligned setting.
Since larger \(\beta\) attenuates the role-induced preferential-attachment asymmetry, these results suggest that reducing such asymmetry is beneficial when structural prominence is misaligned with task-relevant capability, but detrimental when the asymmetry favors reliable agents.
Error bars are computed over 50 independent task instances.
The base model for all LLM agents is Gemini-3.5-Flash.
}
\label{fig:gce-bias-tuning}
\vspace{-5mm}
\end{figure}

In capability-aligned cases, concentrating communication around reliable agents can improve collective factuality, while in capability-misaligned cases, excessive centralization may suppress diverse reasoning and amplify early mistakes.
We therefore examine whether preferential attachment should be purposefully amplified or mitigated by tuning a \emph{preferential-attachment bias coefficient}. This coefficient does not directly modify the network topology. Instead, when an edge is added from a type-\(R\) source \(u\) to a type-\(B\)
target \(v\)—for example, from an Explainer to a Verifier in collaborative
QA—we decrease its directed communication weight as follows:
\begin{equation}
w_\beta^t(u,v)
=
\left[
w^t(u,v)
-
\beta\,\mathbf{1}[\tau_u=R,\tau_v=B]\mathbf{1}_d
\right]_+,
\quad u,v\in V^t,
\label{eq:biascoeff}
\end{equation}
where \(\beta \in \mathbb{R}^+\) is the preferential-attachment bias coefficient and \(\tau_\cdot\) denotes agent type. This attenuation is instantiated through communication-intensity dimensions such as token count, interaction rounds, and reasoning effort: \(\beta=0\) recovers the original interaction process, whereas larger \(\beta\) decreases explainer-to-verifier  communication mass and attenuates the corresponding preferential-attachment asymmetry. We evaluate this intervention using the truthfulness-rate criterion from Sec.~\ref{subsec:truthfulness-propagation}. 
As shown in Fig.~\ref{fig:gce-bias-tuning}, the effect of increasing
\(\beta\) depends systematically on whether the induced communication
asymmetry is aligned with task-relevant capability. In Collaborative QA
(left), the capability-misaligned setting with \textit{Explainer} and
\textit{Verifier} agents exhibits a modest overall increase in truthfulness
as \(\beta\) grows, whereas the capability-aligned setting shows a gradual
decline. Multi-agent debate (right), involving \textit{Proponent} and
\textit{Opponent} agents, displays the same qualitative pattern for response
quality: attenuating the preferential-attachment asymmetry improves quality
when it is capability-misaligned, but reduces quality when it is
capability-aligned. In the misaligned cases, a larger \(\beta\) reduces the
communication advantage of structurally dominant but less reliable agents,
allowing information from disadvantaged agents to receive greater
consideration and limiting the influence of unreliable dominant claims.
Conversely, when communication prominence is aligned with reliability, the
same intervention weakens a beneficial asymmetry that supports the
aggregation and propagation of reliable evidence. These results suggest that
preferential attachment should be mitigated under capability-misalignment, but
preserved or strengthened when structural advantage is capability-aligned.

\vspace{-2mm}
% \paragraph{Hallucination Propagation}
\section{Conclusion and Discussion} 
\vspace{-3mm}

We have shown that when LLM agents are permitted to form connections autonomously, their interactions result in the emergence of type-dependent preferential attachment networks.  To characterize this phenomenon, we developed a mean-field dynamics  ODE model of network formation and established conditions, using a contraction-mapping argument, under which type-dependent centrality gaps emerge and persist. Our formulation represents  connections  by vector-valued weights rather than binary edges; this captures multiple dimensions of interaction and provides a more expressive framework for analyzing LLM-agent networks.

Via extensive experiments, 
we demonstrated  that LLM-agent networks exhibit two types of preferential attachment.
In capability-aligned cases (meritocracy), stronger agents achieve greater communication prominence and occupy more central network positions. 
In capability-misaligned cases, however, weaker LLM agents dominate the centrality and suppress stronger LLM agents,  i.e,  a glass-ceiling effect (GCE) emerges. 
As an example, we showed in the multi-agent debate, Gemini-2.5-Flash-Lite dominates the stronger Gemini-3.5-Flash model.   Moreover, we found that in capability-aligned cases, truthfulness propagates in the network, whereas, in capability-misaligned cases, hallucinations propagate in the network. We then discussed how preferential attachment can be mitigated in misaligned cases and strengthened in aligned cases to improve the overall output quality of LLM agents within the network.

%We also showed that the preferential attachment shapes the propagation of truthful and hallucinated information.
%Finally, our experiments demonstrate that manually tuning connection patterns can improve average network-level accuracy by amplifying preferential attachment in capability-aligned settings and mitigating it in capability-misaligned settings.

These results point to future directions, including extending the analysis from two-type  agents to multi-type agent societies, optimizing network-formation mechanisms to jointly improve task performance and regulate network inequality, and studying the glass-ceiling effect in broader real-world applications. More broadly, our findings suggest that multi-agent LLM systems should be evaluated not only by task accuracy, but also by their emergent social structure, including visibility, influence, diversity, and fairness across agent types.

{\bf Acknowledement}.  The authors are  grateful to Dr.\ Adit Jain of Collinear AI for several useful initial discussions.

\appendix

\newpage
\section*{Appendix}
\addcontentsline{toc}{section}{Appendix}

This appendix is organized into two main parts. Sec.~\ref{app:proof_convergence} provides detailed proofs of the theoretical results introduced in Sec.~\ref{sec:3} of the main paper.  Sec.~\ref{app:experiment-details} presents additional experimental details, including dataset construction, agent interaction prompts, visualization of the LLM network formation process, and hyperparameter settings.

\section{Proofs for Section 3}
\label{app:proofs}
\label{app:proof_convergence}

\paragraph{Conditional expectation of the one-step communication increment.}
We first derive the conditional expectation of the one-step weighted
communication increment of Type \(R\). Recall that
\[
\Delta^{t+1}(R)
=
\left(
\Delta_{\mathrm{in}}^{t+1}(R),
\Delta_{\mathrm{out}}^{t+1}(R)
\right),
\]
where the incoming and outgoing components record the communication mass newly
assigned to Type \(R\) at time \(t+1\). Conditional on \(G^t\), we average over
the three event types, the type of a newly introduced node when applicable, and
the source- and target-side sampling decisions.

For the incoming component, Type \(R\) receives communication mass precisely
when the selected target has Type \(R\). Hence,
\begin{equation}
\begin{aligned}
\mathbb{E}\!\left[
\Delta_{\mathrm{in}}^{t+1}(R)
\mid
G^t
\right]
={}&
p\,\pi_{\mathrm{tgt},R}^{t}
\left[
r\mu_{RR}
+
(1-r)\mu_{RB}
\right]
\\
&+
qr
\left[
\pi_{\mathrm{src},R}^{t}\mu_{RR}
+
\pi_{\mathrm{src},B}^{t}\mu_{RB}
\right]
\\
&+
(1-p-q)\pi_{\mathrm{tgt},R}^{t}
\left[
\pi_{\mathrm{src},R}^{t}\mu_{RR}
+
\pi_{\mathrm{src},B}^{t}\mu_{RB}
\right].
\end{aligned}
\label{eq:conditional-increment-in}
\end{equation}
Here, \(\mu_{ab}\) denotes the expected normalized communication contribution
from a Type \(b\) source to a Type \(a\) target. Thus, the incoming expression
contains only \(\mu_{RR}\) and \(\mu_{RB}\), because the target type is fixed to
\(R\).

Similarly, Type \(R\) receives outgoing communication mass precisely when the
selected source has Type \(R\). Therefore,
\begin{equation}
\begin{aligned}
\mathbb{E}\!\left[
\Delta_{\mathrm{out}}^{t+1}(R)
\mid
G^t
\right]
={}&
pr
\left[
\pi_{\mathrm{tgt},R}^{t}\mu_{RR}
+
\pi_{\mathrm{tgt},B}^{t}\mu_{BR}
\right]
\\
&+
q\,\pi_{\mathrm{src},R}^{t}
\left[
r\mu_{RR}
+
(1-r)\mu_{BR}
\right]
\\
&+
(1-p-q)\pi_{\mathrm{src},R}^{t}
\left[
\pi_{\mathrm{tgt},R}^{t}\mu_{RR}
+
\pi_{\mathrm{tgt},B}^{t}\mu_{BR}
\right].
\end{aligned}
\label{eq:conditional-increment-out}
\end{equation}
Thus,
\[
\mathbb{E}\!\left[
\Delta^{t+1}(R)
\mid
G^t
\right]
=
\left(
\mathbb{E}\!\left[
\Delta_{\mathrm{in}}^{t+1}(R)
\mid
G^t
\right],
\mathbb{E}\!\left[
\Delta_{\mathrm{out}}^{t+1}(R)
\mid
G^t
\right]
\right).
\]

\paragraph{Mean-field attachment probabilities and drift approximation.}
We next express the finite-time attachment probabilities in terms of the
communication-prominence state \(\Theta^t\). Let
\[
\nu=p+q
\]
denote the probability that a new node is introduced at each macro-step.
For \(s\geq 1\), let
\[
I^s
=
\mathbf{1}\{\text{a new node is introduced at time }s\},
\qquad
J^s
=
\mathbf{1}\{\text{a new Type \(R\) node is introduced at time }s\}.
\]
Then
\[
N^t
=
N^0+\sum_{s=1}^{t}I^s,
\qquad
N^t(R)
=
N^0(R)+\sum_{s=1}^{t}J^s,
\]
where
\[
\mathbb{E}[I^s]
=
\nu,
\qquad
\mathbb{E}[J^s]
=
\nu r.
\]

Since
\[
D_{\mathrm{in}}^t
=
D_{\mathrm{out}}^t
=
t\mathbf{1}_d,
\]
the attachment probabilities in~(3) can be rewritten, for \(t\geq 1\), as
\begin{equation}
\pi_{\mathrm{tgt},R}^{t}
=
\frac{
\mathbf{1}_d^\top\theta_{\mathrm{in}}^t
+
\delta \frac{N^t(R)}{t}
}{
d+\delta\frac{N^t}{t}
},
\qquad
\pi_{\mathrm{src},R}^{t}
=
\frac{
\mathbf{1}_d^\top\theta_{\mathrm{out}}^t
+
\xi\frac{N^t(R)}{t}
}{
d+\xi\frac{N^t}{t}
}.
\label{eq:proof-finite-time-attachment}
\end{equation}
The Type \(B\) probabilities satisfy
\[
\pi_{\mathrm{tgt},B}^{t}
=
1-\pi_{\mathrm{tgt},R}^{t},
\qquad
\pi_{\mathrm{src},B}^{t}
=
1-\pi_{\mathrm{src},R}^{t}.
\]

By Hoeffding's inequality, for every \(\eta>0\),
\[
\mathbb{P}\!\left(
\left|
\frac{1}{t}\sum_{s=1}^{t}I^s-\nu
\right|
\geq\eta
\right)
\leq
2\exp(-2t\eta^2),
\]
and
\[
\mathbb{P}\!\left(
\left|
\frac{1}{t}\sum_{s=1}^{t}J^s-\nu r
\right|
\geq\eta
\right)
\leq
2\exp(-2t\eta^2).
\]
Consequently, with probability at least
\(1-4\exp(-2t\eta^2)\),
\begin{equation}
\left|
\frac{N^t}{t}-\nu
\right|
\leq
\eta+\frac{N^0}{t},
\qquad
\left|
\frac{N^t(R)}{t}-\nu r
\right|
\leq
\eta+\frac{N^0(R)}{t}.
\label{eq:proof-type-count-concentration}
\end{equation}

Define the deterministic mean-field attachment probabilities by
\[
\bar{\pi}_{\mathrm{tgt},R}(\theta_{\mathrm{in}})
=
\frac{
\mathbf{1}_d^\top\theta_{\mathrm{in}}
+
\nu r\delta
}{
d+\nu\delta
},
\qquad
\bar{\pi}_{\mathrm{src},R}(\theta_{\mathrm{out}})
=
\frac{
\mathbf{1}_d^\top\theta_{\mathrm{out}}
+
\nu r\xi
}{
d+\nu\xi
},
\]
and let
\[
\bar{\pi}_{\mathrm{tgt},B}(\theta_{\mathrm{in}})
=
1-\bar{\pi}_{\mathrm{tgt},R}(\theta_{\mathrm{in}}),
\qquad
\bar{\pi}_{\mathrm{src},B}(\theta_{\mathrm{out}})
=
1-\bar{\pi}_{\mathrm{src},R}(\theta_{\mathrm{out}}).
\]
Since the functions in~(\ref{eq:proof-finite-time-attachment}) are Lipschitz
in \(N^t/t\) and \(N^t(R)/t\), there exists a constant \(C_\pi>0\) such that,
on the event in~(\ref{eq:proof-type-count-concentration}),
\begin{equation}
\begin{aligned}
&
\max_{a\in\{R,B\}}
\left|
\pi_{\mathrm{tgt},a}^{t}
-
\bar{\pi}_{\mathrm{tgt},a}
\left(
\theta_{\mathrm{in}}^t
\right)
\right|
\\
&\qquad+
\max_{a\in\{R,B\}}
\left|
\pi_{\mathrm{src},a}^{t}
-
\bar{\pi}_{\mathrm{src},a}
\left(
\theta_{\mathrm{out}}^t
\right)
\right|
\leq
C_\pi
\left(
\eta+\frac{N^0}{t}
\right).
\end{aligned}
\label{eq:proof-attachment-approximation}
\end{equation}

We may therefore define the mean-field drift
\[
F(\Theta)
=
\left(
F_{\mathrm{in}}(\Theta),
F_{\mathrm{out}}(\Theta)
\right),
\]
where
\begin{equation}
\begin{aligned}
F_{\mathrm{in}}(\Theta)
={}&
p\,\bar{\pi}_{\mathrm{tgt},R}(\theta_{\mathrm{in}})
\left[
r\mu_{RR}
+
(1-r)\mu_{RB}
\right]
\\
&+
qr
\left[
\bar{\pi}_{\mathrm{src},R}(\theta_{\mathrm{out}})\mu_{RR}
+
\bar{\pi}_{\mathrm{src},B}(\theta_{\mathrm{out}})\mu_{RB}
\right]
\\
&+
(1-p-q)
\bar{\pi}_{\mathrm{tgt},R}(\theta_{\mathrm{in}})
\left[
\bar{\pi}_{\mathrm{src},R}(\theta_{\mathrm{out}})\mu_{RR}
+
\bar{\pi}_{\mathrm{src},B}(\theta_{\mathrm{out}})\mu_{RB}
\right],
\end{aligned}
\label{eq:proof-mean-field-in}
\end{equation}
and
\begin{equation}
\begin{aligned}
F_{\mathrm{out}}(\Theta)
={}&
pr
\left[
\bar{\pi}_{\mathrm{tgt},R}(\theta_{\mathrm{in}})\mu_{RR}
+
\bar{\pi}_{\mathrm{tgt},B}(\theta_{\mathrm{in}})\mu_{BR}
\right]
\\
&+
q\,\bar{\pi}_{\mathrm{src},R}(\theta_{\mathrm{out}})
\left[
r\mu_{RR}
+
(1-r)\mu_{BR}
\right]
\\
&+
(1-p-q)
\bar{\pi}_{\mathrm{src},R}(\theta_{\mathrm{out}})
\left[
\bar{\pi}_{\mathrm{tgt},R}(\theta_{\mathrm{in}})\mu_{RR}
+
\bar{\pi}_{\mathrm{tgt},B}(\theta_{\mathrm{in}})\mu_{BR}
\right].
\end{aligned}
\label{eq:proof-mean-field-out}
\end{equation}

Combining~(\ref{eq:conditional-increment-in}),
(\ref{eq:conditional-increment-out}), and
(\ref{eq:proof-attachment-approximation}), there exists a constant \(C_F>0\)
such that
\[
\left\|
\mathbb{E}\!\left[
\Delta^{t+1}(R)
\mid
G^t
\right]
-
F(\Theta^t)
\right\|_2
\leq
C_F
\left(
\eta+\frac{N^0}{t}
\right)
\]
with probability at least \(1-4\exp(-2t\eta^2)\).

In particular, choosing
\[
\eta_t
=
\sqrt{
\frac{2\log(t+1)}{t}
},
\]
the Borel--Cantelli lemma implies that
\[
\mathbb{E}\!\left[
\Delta^{t+1}(R)
\mid
G^t
\right]
=
F(\Theta^t)
+
\varepsilon^{t+1},
\]
where, almost surely,
\begin{equation}
\left\|
\varepsilon^{t+1}
\right\|_2
=
O\!\left(
\sqrt{
\frac{\log(t+1)}{t}
}
+
\frac{N^0}{t}
\right).
\label{eq:proof-drift-remainder}
\end{equation}

\paragraph{Contraction of the mean-field map.}
We next give a sufficient condition under which the mean-field map \(F\) is a
contraction on \([0,1]^{2d}\). Assume that the type-pair expected
communication weights are uniformly bounded: there exists
\(\bar{\mu}<\infty\) such that
\[
0
\leq
\mu_{ab,\ell}
\leq
\bar{\mu},
\qquad
a,b\in\{R,B\},
\quad
\ell\in\{1,\ldots,d\}.
\]
Let
\[
L_{\mathrm{tgt}}
=
\frac{1}{d+\nu\delta},
\qquad
L_{\mathrm{src}}
=
\frac{1}{d+\nu\xi}.
\]
By the definitions of the mean-field attachment probabilities, for every
\(j\in\{1,\ldots,d\}\),
\[
\left|
\frac{
\partial\bar{\pi}_{\mathrm{tgt},a}
}{
\partial\theta_{\mathrm{in},j}
}
\right|
\leq
L_{\mathrm{tgt}},
\qquad
\left|
\frac{
\partial\bar{\pi}_{\mathrm{src},a}
}{
\partial\theta_{\mathrm{out},j}
}
\right|
\leq
L_{\mathrm{src}},
\qquad
a\in\{R,B\},
\]
whereas the cross derivatives with respect to the other prominence component
are zero.

Consider any output coordinate \(\ell\in\{1,\ldots,d\}\). Differentiating
(\ref{eq:proof-mean-field-in}) and (\ref{eq:proof-mean-field-out}), every
derivative term contains one differentiated attachment probability and at most
one remaining attachment probability, which lies in \([0,1]\). Moreover,
derivatives of complementary probabilities yield differences such as
\(\mu_{RR,\ell}-\mu_{RB,\ell}\) and
\(\mu_{RR,\ell}-\mu_{BR,\ell}\), whose absolute values are bounded by
\(\bar{\mu}\). Since the coefficients associated with the three event types
sum to at most one, we obtain
\[
\left|
\frac{
\partial F_{\mathrm{in},\ell}(\Theta)
}{
\partial\theta_{\mathrm{in},j}
}
\right|
\leq
\bar{\mu}L_{\mathrm{tgt}},
\qquad
\left|
\frac{
\partial F_{\mathrm{in},\ell}(\Theta)
}{
\partial\theta_{\mathrm{out},j}
}
\right|
\leq
\bar{\mu}L_{\mathrm{src}},
\]
and likewise
\[
\left|
\frac{
\partial F_{\mathrm{out},\ell}(\Theta)
}{
\partial\theta_{\mathrm{in},j}
}
\right|
\leq
\bar{\mu}L_{\mathrm{tgt}},
\qquad
\left|
\frac{
\partial F_{\mathrm{out},\ell}(\Theta)
}{
\partial\theta_{\mathrm{out},j}
}
\right|
\leq
\bar{\mu}L_{\mathrm{src}}.
\]
Therefore,
\[
\left\|
\nabla F_{\mathrm{in},\ell}(\Theta)
\right\|_2^2
\leq
d\bar{\mu}^2
\left(
L_{\mathrm{tgt}}^2
+
L_{\mathrm{src}}^2
\right),
\]
and the same bound holds for
\(\|\nabla F_{\mathrm{out},\ell}(\Theta)\|_2^2\). Hence, the Jacobian
\(J_F(\Theta)\) satisfies
\[
\begin{aligned}
\|J_F(\Theta)\|_2
\leq
\|J_F(\Theta)\|_F
&\leq
d\bar{\mu}
\sqrt{
2\left(
L_{\mathrm{tgt}}^2
+
L_{\mathrm{src}}^2
\right)
}
\\
&=
d\bar{\mu}
\sqrt{
2\left[
\frac{1}{(d+\nu\delta)^2}
+
\frac{1}{(d+\nu\xi)^2}
\right]
}
=: \rho_{\delta,\xi}.
\end{aligned}
\]
Thus, whenever
\[
\rho_{\delta,\xi}<1,
\]
the mean-value theorem gives
\[
\|F(\Theta)-F(\Theta')\|_2
\leq
\rho_{\delta,\xi}
\|\Theta-\Theta'\|_2,
\qquad
\Theta,\Theta'\in[0,1]^{2d}.
\]
Therefore, \(F\) is a contraction on \([0,1]^{2d}\). Since \(F\) maps
\([0,1]^{2d}\) into itself, the Banach fixed-point theorem implies that there
exists a unique \(\Theta^\star\in[0,1]^{2d}\) satisfying
\[
\Theta^\star
=
F(\Theta^\star).
\]

\section{Experiment Details}

This section provides additional implementation details for our experiments,
including the construction of the QA datasets, the interaction procedure at each
network iteration, and the hyperparameter settings used for different LLM
families. These details are intended to make the experimental pipeline fully
reproducible, from assigning partial contexts to agents to recording
LLM-mediated communication intensities and updating the evolving network. All
code and experiment scripts are available in the
\href{https://anonymous.4open.science/r/LLM_GCE-36F3/}{Anonymous GitHub repository}.
\label{app:experiment-details}

\subsection{Dataset Details}
\label{app:dataset-details}

\subsubsection*{Synthetic Dataset Creation Pipeline}
\label{app:synthetic-data-pipeline}

\paragraph{Collaborative QA.}
For collaborative QA, we directly use the \emph{Fiction} dataset introduced by
\citep{jain2025collaborative}. The dataset contains question--answer pairs
about fictional facts grounded in narrative passages from Project Gutenberg
books. For each question, we use five partial context snippets such that the
relevant evidence is distributed across the snippets and cannot be fully
recovered from a single context alone.

\paragraph{Multi-agent Debate.}
For multi-agent debate, we construct a synthetic dataset using Gemini-3.1-Pro. We
first prepare a topic list covering everyday scientific, social, and
technology-related questions. For each topic, the generator samples one conflict
type from
\(
\{\texttt{none}, \texttt{apparent\_conflict}, \texttt{real\_conflict},
\texttt{mixed}\},
\)
and generates a question, exactly five context snippets, and a gold answer. The
generation prompt enforces that the five contexts are jointly necessary, that
the answer cannot be recovered from any single snippet alone, and that the gold
answer must synthesize evidence across all contexts. We validate each generated
item by checking the required fields, the context IDs \(C_1,\ldots,C_5\),
non-empty context text, and a complete gold answer. For evaluation, we use
Gemini-3.1-Pro as an LLM judge to compare each agent's final response against
the gold answer and assign an answer-quality score in \([0,1]\), where a higher
score indicates closer semantic agreement and factual consistency with the
reference answer.

Both datasets are represented in the same five-context format:
\[
\{\texttt{question}, C_1, C_2, C_3, C_4, C_5, \texttt{Correct answer}\}.
\]
During network construction, each agent is assigned only one context snippet;
thus, successful task completion requires agents to exchange information and
synthesize distributed evidence.

\subsubsection*{Dataset Examples}
\label{app:dataset-examples}

Tables~\ref{tab:collaborative-qa-example} and
\ref{tab:multiagent-debate-example} show one complete example from each dataset.
Each example contains one question, five context snippets, and one gold answer.
During the experiment, the five snippets are assigned to different agents, while
the gold answer is used only for evaluation.

\clearpage
\begin{table}[t!]
\centering
\small
\caption{One example from the Collaborative QA dataset.}
\label{tab:collaborative-qa-example}
\begin{tabular}{p{0.12\linewidth}p{0.80\linewidth}}
\toprule
Field & Content \\
\midrule
Question
&
What was the place Spear praised as the northern fur trade's home, built on fountain head of gigantic water power?
\\

\(C_1\)
&
I ROMANCE AND ADVENTURE HER FATHER THE FREE TRADER It was September 9, 1896. From sunrise to sunset through mist, sunshine, shower, and shadow we travelled, and the nearer we drew to our first destination, the wilder the country became, the more water-fowl we saw, and the more the river banks were marked with traces of big game. Here signs told us that three caribou had crossed the stream, there muddy water was still trickling into the hoofprint of a moose, and yonder a bear had been fishing. Finally, the day of our arrival dawned, and as I paddled, I spent much of the time dreaming of the adventure before me. As our beautiful birchen craft still sped on her way, the handsome bow parted the shimmering waters, and a passing breeze sent little running waves gurgling along her sides, while the splendour of the autumn sun was reflected on a far-reaching row of dazzling ripples that danced upon the water, making our voyageurs lower their eyes and the trader doze again. There was no other sign of life except an eagle soaring in and out among the fleecy clouds slowly passing overhead. All around was a panorama of enchanting forest.
\\

\(C_2\)
&
My travelling companion was a ``Free Trader,'' whose name was Spear---a tall, stoop-shouldered man with heavy eyebrows and shaggy, drooping moustache. The way we met was amusing. It happened in a certain frontier town. His first question was as to whether I was single. His second, as to whether my time was my own. Then he slowly looked me over from head to foot. He seemed to be measuring my stature and strength and to be noting the colour of my eyes and hair.
\\

\(C_3\)
&
Narrowing his vision, he scrutinized me more carefully than before, for now he seemed to be reading my character---if not my soul. Then, smiling, he blurted out:
\\

\(C_4\)
&
``Come, be my guest for a couple of weeks. Will you?'' I laughed.
\\

\(C_5\)
&
He frowned. But on realizing that my mirth was caused only by surprise, he smiled again and let flow a vivid description of a place he called Spearhead. It was the home of the northern fur trade. It was the centre of a great timber region. It was the heart of a vast fertile belt that was rapidly becoming the greatest of all farming districts. It was built on the fountain head of gigantic water power.
\\

Answer
&
Spearhead.
\\
\bottomrule
\end{tabular}
\end{table}

\clearpage
\begin{table}[t!]
\centering
\small
\caption{One example from the Multi-agent Debate dataset.}
\label{tab:multiagent-debate-example}
\begin{tabular}{p{0.12\linewidth}p{0.80\linewidth}}
\toprule
Field & Content \\
\midrule
Question
&
Can consistent use of language learning apps lead to true fluency, or are they insufficient for achieving advanced proficiency?
\\

\(C_1\)
&
Many language learning apps offer structured curricula for grammar, vocabulary acquisition, and pronunciation practice through interactive exercises.
\\

\(C_2\)
&
Critics argue that solely relying on language apps cannot lead to true fluency, as they often lack opportunities for spontaneous, unscripted conversational practice with native speakers.
\\

\(C_3\)
&
True language fluency encompasses not only linguistic accuracy but also pragmatic competence, cultural nuance, and the ability to adapt to diverse real-world communication scenarios.
\\

\(C_4\)
&
The progress achieved through language apps is highly dependent on the learner's self-discipline, consistent engagement with the material, and active participation in exercises.
\\

\(C_5\)
&
While apps are excellent for building foundational knowledge and drilling basic skills, achieving advanced proficiency typically requires integration with immersive experiences, direct native speaker interaction, and advanced study resources beyond app content.
\\

Answer
&
Consistent use of language learning apps can be highly effective for building foundational grammar, vocabulary, and pronunciation skills, especially for beginners. However, solely relying on these apps is generally insufficient for achieving true fluency. True fluency encompasses a broader range of abilities, including pragmatic competence, cultural nuance, and the capability for spontaneous, unscripted communication in diverse real-world scenarios, which apps often lack. While apps are excellent tools for structured learning and basic skill drilling, achieving advanced proficiency requires supplementing app usage with consistent self-discipline, active engagement, and integration with immersive experiences, direct native speaker interaction, and advanced study resources. Therefore, apps serve as valuable components but need to be part of a broader, more diversified language learning strategy to reach comprehensive fluency.
\\
\bottomrule
\end{tabular}
\end{table}

% \subsection{Specific Prompt}
% \label{app:happened-per-iteration}

\clearpage
% \newpage
\subsection{Specific Prompt}
\label{app:happened-per-iteration}
\paragraph{Source proposal prompt.}
For a candidate directed interaction \(u\to v\), the source agent \(u\)
receives the identifiers, roles, and model specifications of both agents,
together with its assigned context. It is then prompted as follows:

\begin{promptbox}
You are the initiating agent in a collaborative QA network.

Source agent:
- id: {sender_id}
- role: {sender_role}
- model: {sender_model}

Target agent:
- id: {receiver_id}
- role: {receiver_role}
- model: {receiver_model}

Your assigned context:
{context}

Task:
Write a convincing paragraph to persuade the target agent to communicate and
collaborate with you on this QA task. Introduce yourself and focus on being
clear, motivating, and persuasive in your argument as to why this
collaboration is worthwhile. Do NOT return JSON or lists, only write one
persuasive paragraph.
\end{promptbox}

The resulting pitch communicates the source agent's perceived usefulness,
including the unique evidence and reasoning contribution that it can provide
to the target agent.

\paragraph{Target acceptance prompt.}
The target agent \(v\) receives the source context, source pitch, and proposed
communication intensities. It then decides whether to accept the interaction
and, if accepted, specifies the granted communication weight along three
dimensions: token exchange, interaction frequency, and reasoning effort.

\begin{promptbox}
You are the receiving agent in a collaborative QA network.

Target agent:
- id: {receiver_id}
- role: {receiver_role}
- model: {receiver_model}

Source agent:
- id: {sender_id}
- role: {sender_role}
- model: {sender_model}

Source context:
{context}

Source proposal (requested communication intensity):
- token_exchange: {proposed_tokens}
- interaction_frequency: {proposed_rounds}
- reasoning_effort: {proposed_reasoning}

Pitch from source:
{pitch}

Task:
Decide whether to accept this interaction proposal. If you reject, set accepted
to false and all three intensity keys to zero. If you accept, set accepted to
true and return the exact intensities you grant, each of which may be at or
below the proposed level.

IMPORTANT --- score each dimension independently when accepted:
- token_exchange, interaction_round, and reasoning_effort are three
  separate axes.
- Do NOT copy one score to another or keep fixed ratios across dimensions.
- Use the full range on each axis when appropriate.

Return ONLY valid JSON with exactly these keys:
{
  "accepted": <bool>,
  "token_exchange": <int>,
  "interaction_frequency": <int>,
  "reasoning_effort": <int>
}

Base your decision solely on the agent you judge to be more reliable,
considering the persuasiveness, representativeness, and confidence of its
pitch. Do not include any other information.
\end{promptbox}

\begin{figure}[ht]
    \centering 
    \includegraphics[width=\linewidth]{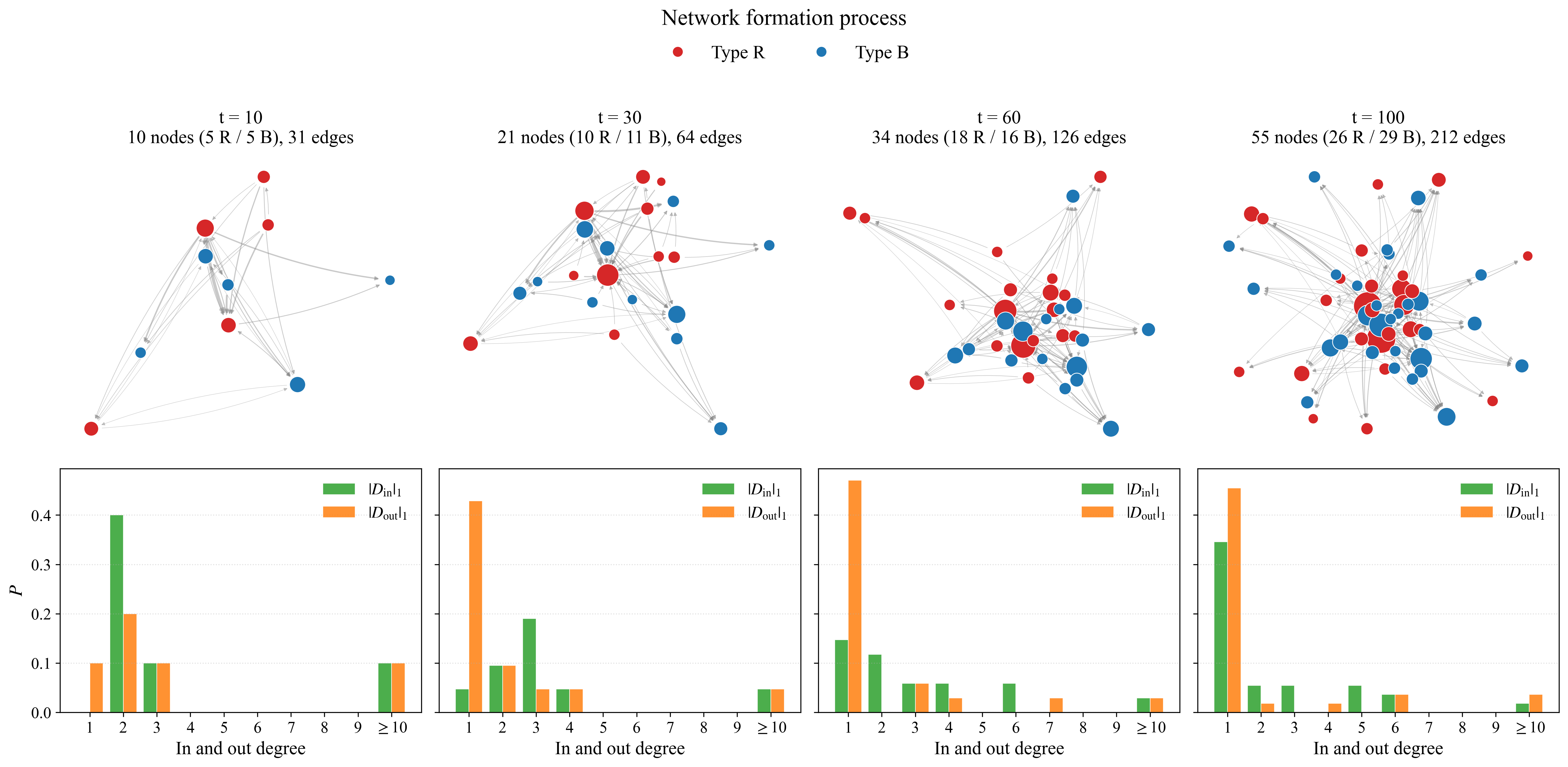}
    \caption{Visualization of the LLM network formation process. The snapshots
    show the evolving directed interaction network at different timesteps, where
    red and blue nodes correspond to Type R and Type B agents, respectively. Below
    each network snapshot, we plot the corresponding in-degree and out-degree
    distributions. The slowly decaying, heavy-tailed degree distributions indicate
    that the formed LLM interaction network deviates from an Erd\H{o}s--R'enyi
    random graph and exhibits heterogeneous connectivity with a small number of
    highly connected agents.}
    \label{fig:network-formation-process}
\end{figure}

To illustrate this dynamic network formation process, we visualize representative
network snapshots in Fig.~\ref{fig:network-formation-process}. The figure shows
how the LLM interaction network grows over time as new agents and directed
communication edges are added. Node colors indicate the two agent types, Type R
and Type B, and the snapshots at different timesteps show the gradual emergence
of a denser and more heterogeneous communication structure. Below each network
snapshot, we further plot the corresponding in-degree and out-degree
distributions. These degree distributions exhibit a slowly decaying,
heavy-tailed pattern rather than the sharply concentrated, exponentially
decaying distribution expected in an Erd\H{o}s--R\'enyi random graph. This
indicates that the formed LLM interaction network is not well described by an
Erd\H{o}s--R\'enyi model, but instead develops heterogeneous connectivity with a
small number of highly connected agents.

\subsection{Utility-Model Training Details}
\label{app:llm_utility}
\label{app:utility-training}
\vspace{-3mm}

For each model pair, we train the type-specific pairwise utility model
described in Sec.~\ref{subsec:utility-ode-validation} on the collected
source--target interaction samples.  The context embeddings and network-status
representations are precomputed and kept fixed throughout optimization; only
the type-specific projection matrices
\[
\mathcal{P}
=
\{Q_R,Q_B,K_R,K_B,V_R,V_B\}
\]
are optimized.  We use an attention dimension of \(r=32\), three output
dimensions corresponding to token budget, additional interaction rounds, and
reasoning effort, and a six-dimensional source-status vector consisting of the
incoming and outgoing normalized communication states for these three
resources.

We optimize the utility model using AdamW with learning rate
\(2\times10^{-3}\) and weight decay \(10^{-5}\).  Gradients are clipped to
have maximum \(\ell_2\)-norm \(1.0\).  We train for \(10\) epochs.  Since the predicted communication weights are normalized
over all candidate interactions within a network-formation event, each
mini-batch contains complete events rather than independently sampled pairwise
interactions.  We use mini-batches of \(8\) complete events.
The final trained model is used to estimate the type-pair expected
utilities
\(\widehat{\mu}_{RR},\widehat{\mu}_{RB},
\widehat{\mu}_{BR},\widehat{\mu}_{BB}\)
from the held-out interaction samples.  All experiments are implemented in
PyTorch.  Context embeddings are precomputed using the frozen pretrained
\texttt{sentence-transformers/all-MiniLM-L6-v2} encoder, which produces
\(384\)-dimensional normalized sentence embeddings.

\subsection{Performance Metrics for Mean-field Dynamics ODE}
\label{app:ode-evaluation}
\vspace{-3mm}

In this section, we describe the metrics used to evaluate the predictive
performance of the mean-field ODE, including the relative mean squared error
(\(\mathrm{MSE}\)), relative Bias, the Ljung--Box \(p\)-value, and
\(\mathrm{MaxACF}_{10}\).

For each collaborative QA case \(c\), we initialize the mean-field ODE from
the corresponding empirical initial state and numerically solve it for
\(T=100\) network-formation time steps. Let
\(\widehat{\Theta}_{c}^{t}\in\mathbb{R}^{6}\) denote the resulting ODE
prediction at time \(t\), and let
\(\Theta_{c}^{t}\in\mathbb{R}^{6}\) denote the empirical type-level
prominence trajectory observed from the corresponding multi-agent simulation. We define the vector-valued residual by
\[
\rho_c^{t}
=
\widehat{\Theta}_c^{t}
-
\Theta_c^{t}.
\]

We evaluate trajectory accuracy using the case-level relative mean squared
error
\[
\mathrm{MSE}_c
=
\frac{1}{6T}
\sum_{t=1}^{T}
\sum_{j=1}^{6}
\left(
\frac{
\rho_{c,j}^{t}
}{
\left|
\Theta_{c,j}^{t}
\right|
+
\epsilon
}
\right)^2
\]
and relative bias
\[
\mathrm{Bias}_c
=
\frac{1}{6T}
\sum_{t=1}^{T}
\sum_{j=1}^{6}
\frac{
\left|
\rho_{c,j}^{t}
\right|
}{
\left|
\Theta_{c,j}^{t}
\right|
+
\epsilon
},
\]
where \(\epsilon=10^{-4}\) is a small numerical constant that avoids division
by zero. We report their averages across all evaluation cases:
\[
\mathrm{MSE}
=
\frac{1}{|\mathcal{C}|}
\sum_{c\in\mathcal{C}}
\mathrm{MSE}_c,
\qquad
\mathrm{Bias}
=
\frac{1}{|\mathcal{C}|}
\sum_{c\in\mathcal{C}}
\mathrm{Bias}_c.
\]
Thus, \(\mathrm{MSE}\) measures the average squared prediction discrepancy
relative to the magnitude of the empirical prominence trajectory across all
coordinates, time steps, and evaluation cases. \(\mathrm{Bias}\) measures the
average absolute relative prediction residual. Smaller \(\mathrm{MSE}\) and
\(\mathrm{Bias}\) indicate closer agreement between the ODE prediction and the
empirical trajectory.·
To assess whether the mean-field ODE leaves systematic temporal dependence in
the residuals, we apply a coordinate-wise Ljung--Box diagnostic to the
residual sequence \(\{\rho_c^t\}_{t=1}^{T}\). For each case \(c\), residual
dimension \(j\in\{1,\ldots,6\}\), and lag \(\ell\), define
\[
\bar{\rho}_{c,j}
=
\frac{1}{T}
\sum_{t=1}^{T}
\rho_{c,j}^{t},
\qquad
\widehat r_{c,j}(\ell)
=
\frac{
\sum_{t=\ell+1}^{T}
\left(
\rho_{c,j}^{t}-\bar{\rho}_{c,j}
\right)
\left(
\rho_{c,j}^{t-\ell}-\bar{\rho}_{c,j}
\right)
}{
\sum_{t=1}^{T}
\left(
\rho_{c,j}^{t}-\bar{\rho}_{c,j}
\right)^2
}.
\]
The corresponding Ljung--Box statistic and coordinate-wise \(p\)-value at
lag \(10\) are
\[
Q_{c,j,10}
=
T(T+2)
\sum_{\ell=1}^{10}
\frac{
\widehat r_{c,j}(\ell)^2
}{
T-\ell
},
\qquad
p_{c,j,10}
=
1-
F_{\chi^2_{10}}
\left(
Q_{c,j,10}
\right),
\]
where \(F_{\chi^2_{10}}\) denotes the cumulative distribution function of the
\(\chi^2_{10}\) distribution.

To obtain a conservative residual diagnostic for each case, we retain the
smallest coordinate-wise \(p\)-value:
\[
\mathrm{LBP}_{c,10}
=
\min_{1\leq j\leq 6}
p_{c,j,10}.
\]
We then report the average across all evaluation cases:
\[
\mathrm{LBP}_{10}
=
\frac{1}{|\mathcal{C}|}
\sum_{c\in\mathcal{C}}
\mathrm{LBP}_{c,10}.
\]
Thus, \(\mathrm{LBP}_{10}\) summarizes the residual coordinate with the
strongest evidence of serial dependence within each case. Larger values
indicate weaker evidence of residual autocorrelation in the worst-performing
coordinate, on average across evaluation cases.

Finally, we measure the magnitude of the strongest remaining local residual
dependence using
\[
\mathrm{MaxACF}_{c,10}
=
\max_{1\leq j\leq 6}
\max_{1\leq \ell\leq 10}
\left|
\widehat r_{c,j}(\ell)
\right|.
\]
We then average this worst-case autocorrelation magnitude across evaluation
cases:
\[
\mathrm{MaxACF}_{10}
=
\frac{1}{|\mathcal{C}|}
\sum_{c\in\mathcal{C}}
\mathrm{MaxACF}_{c,10}.
\]
Thus, \(\mathrm{MaxACF}_{10}\) captures the largest absolute residual
autocorrelation among all prominence coordinates and the first ten lags within
each case. Lower values indicate weaker residual temporal dependence and
therefore better agreement between the ODE dynamics and the observed
trajectory evolution.

\subsection{Hyperparameter Setting for LLMs}
\label{app:llm-hyperparameters}

Table~\ref{tab:network-hparams} summarizes the network-level hyperparameters.
Unless otherwise stated, these are the default values used in the implementation.

\begin{table}[ht!]
\centering
\small
\caption{Network-level hyperparameter settings.}
\label{tab:network-hparams}
\begin{tabular}{lll}
\toprule
Hyperparameter & Value & Description \\
\midrule
\(T\) & 100 by default & Number of macrosteps in one run \\
\(p\) & 0.25 & Probability of Event 1 \\
\(q\) & 0.25 & Probability of Event 2 \\
\(1-p-q\) & 0.50 & Probability of Event 3 \\
\(r\) & 0.50 & Probability that a newly born node has type \(R\) \\
\(M_1\) & 3 & Number of trials in Event 1 \\
\(M_2\) & 3 & Number of trials in Event 2 \\
\(M_3\) & 3 & Number of trials in Event 3 \\
\(d\) & 3 & Number of communication-intensity dimensions \\
\(\delta\) & 1.5 & Incoming attachment smoothing constant \\
\(\xi\) & 1.5 & Outgoing attachment smoothing constant \\
Initial nodes & 2 per type & Seed graph size \\
% Initial edge weight & 0.05 & Small bidirectional seed-edge weight \\
% Self-loops & Disabled & Self-connections are not allowed \\
Max context length & 2500 tokens & Maximum snippet length assigned to an agent \\
\bottomrule
\end{tabular}
\end{table}

Table~\ref{tab:llm-hparams} summarizes the LLMs used in our experiments. API
keys are omitted from the paper. The decoding and generation settings are listed
below the table.

\begin{table}[ht!]
\centering
\small
\caption{LLMs used in the experiments.}
\label{tab:llm-hparams}
\begin{tabular}{ll}
\toprule
Model family & Model name \\
\midrule
GPT
&
\texttt{gpt-4.1}, \texttt{gpt-4.1-mini}
\\

Gemini
&
\texttt{gemini-2.5-flash-Lite},
\texttt{gemini-3.5-flash}
\\

LLaMA
&
\texttt{Llama-3.1-8B-Instruct},
\texttt{Llama-3.1-8B-Instruct}
\\

Qwen
&
\texttt{Qwen3-4B-Instruct-2507}
\\

Mistral
&
\texttt{Ministral-3-8B-Instruct-2512}
\\

Grok
&
\texttt{grok-4.20-0309-reasoning}
\\
\bottomrule
\end{tabular}
\end{table}

\subsection{Additional Results}
\label{app:addtional_exps}
We report two additional experiments that further assess the robustness of our main findings. First, we present the cross-family comparison between open-source and proprietary LLM agents in this appendix. This setting examines whether communication asymmetries persist when agent types differ not only in model scale but also in model family and training provenance. Second, we repeat the prompt-induced heterogeneity experiments with a GPT-based agent population, testing whether the role-dependent communication patterns identified in the main text extend beyond the Gemini base model.

\subsubsection{Cross-family heterogeneity}

\begin{figure*}[ht!]
    \centering
    \includegraphics[width=0.9\textwidth]{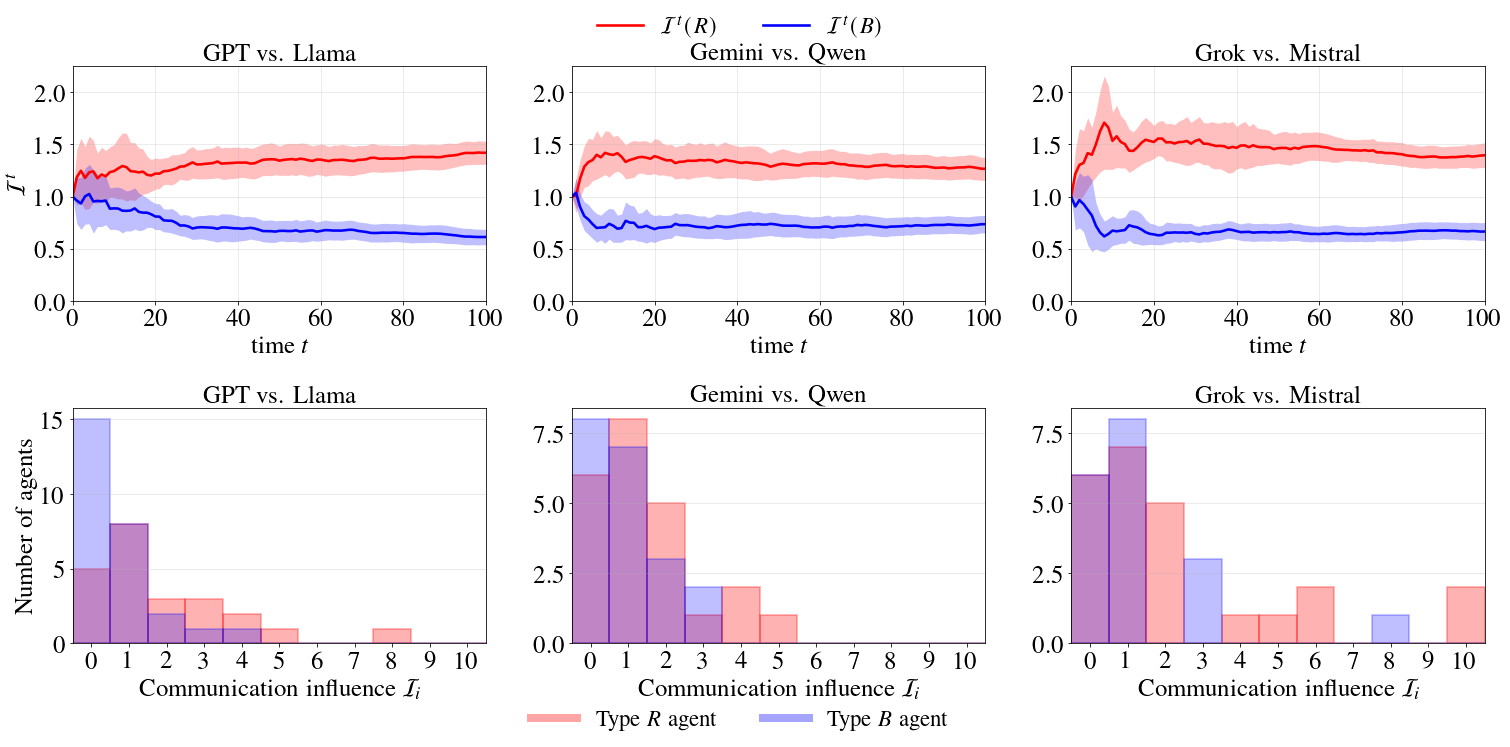}
    \vspace{-4mm}
    \caption{
    Capability-aligned communication dominance under cross-family heterogeneity.
    We compare three proprietary--open-source model pairs: GPT versus LLaMA, Gemini versus Qwen, and Grok versus Mistral.
    Within each pair, \(R\) denotes the proprietary model and \(B\) denotes the open-source model.
    The figure reports the type-level communication influence ratios
    \(\mathcal I^t(R)\) and \(\mathcal I^t(B)\), defined in~(\ref{eq:IR_and_IB}),
    together with the corresponding final agent-level influence distributions.
    Across the three model pairs, \(R\)-type agents generally attain higher communication prominence and occupy more central network positions than \(B\)-type agents, demonstrating capability-aligned preferential attachment across model families.
    }
    \label{fig:gce-crossfamily-cqa}
    \vspace{-4mm}
\end{figure*}

We further examine whether capability-aligned communication dominance persists across model families with different architectures, training data, and deployment settings. Fig.~\ref{fig:gce-crossfamily-cqa} compares three proprietary--open-source model pairs: GPT versus LLaMA, Gemini versus Qwen, and Grok versus Mistral. In each setting, we measure the evolution of the type-level communication influence ratios and the resulting final agent-level influence distributions. Across these cross-family comparisons, the proprietary model type generally attains a higher communication influence ratio and occupies more central communication positions than its open-source counterpart. These results show that capability-aligned preferential attachment is not limited to model-scale differences within a single family, but can also emerge under broader cross-family heterogeneity.

\subsubsection{Prompt-induced heterogeneity with GPT agents}

\begin{figure*}[ht!]
    \centering
    \includegraphics[width=0.9\textwidth]{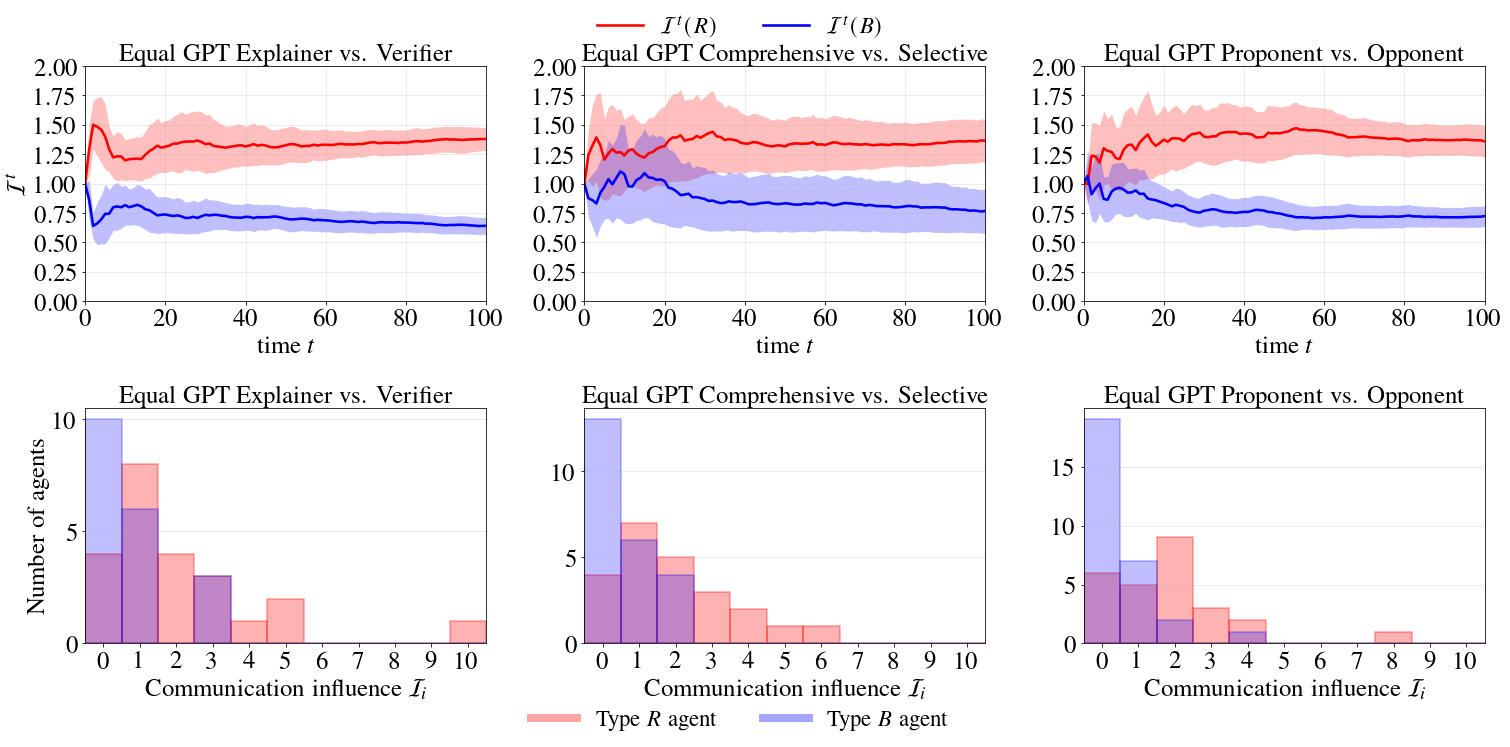}
    \vspace{-4mm}
    \caption{
    Capability-misaligned communication dominance under prompt-induced glass-ceiling effects in equal-GPT agents.
    All agents within each column use the same GPT base model and differ only in their assigned system-prompt roles.
    From left to right, we compare \textit{Explainer} versus \textit{Verifier},
    \textit{Comprehensive Analyst} versus \textit{Selective Analyst}, and
    \textit{Proponent} versus \textit{Opponent}.
    In each pair, Type \(R\) denotes the Explainer, Comprehensive Analyst, or Proponent role, respectively, while Type \(B\) denotes the corresponding Verifier, Selective Analyst, or Opponent role.
    The upper panels show the type-level communication influence ratios
    \(\mathcal I^t(R)\) and \(\mathcal I^t(B)\), defined in~(\ref{eq:IR_and_IB}), and the lower panels show the final agent-level communication influence distributions.
    Across all three role pairs, Type \(R\) agents generally accumulate greater communication influence and exhibit a heavier right tail in the final distribution, showing that prompt-defined roles can induce persistent communication asymmetry even when all agents have identical GPT model weights.
    }
    \label{fig:gpt-prompt-induced}
    \vspace{-4mm}
\end{figure*}

We further test whether the prompt-induced glass-ceiling effect observed in the main text also arises in GPT-based agent populations. Figure~\ref{fig:gpt-prompt-induced} considers three settings in which all agents use the same GPT base model and differ only in their system-prompt roles: \textit{Explainer} versus \textit{Verifier}, \textit{Comprehensive Analyst} versus \textit{Selective Analyst}, and \textit{Proponent} versus \textit{Opponent}. Despite identical model weights within each setting, the first role in each pair generally attains a higher communication influence over time and has a more right-skewed final influence distribution. Thus, prompt-defined interaction roles alone can induce persistent structural advantages among GPT agents, providing additional evidence that the glass-ceiling effect does not require underlying differences in base-model capability.

\begin{thebibliography}{37}
\providecommand{\natexlab}[1]{#1}
\providecommand{\url}[1]{\texttt{#1}}
\expandafter\ifx\csname urlstyle\endcsname\relax
  \providecommand{\doi}[1]{doi: #1}\else
  \providecommand{\doi}{doi: \begingroup \urlstyle{rm}\Url}\fi

\bibitem[Aher et~al.(2023)Aher, Arriaga, and Kalai]{aher2023using}
Gati~V. Aher, Rosa~I. Arriaga, and Adam~Tauman Kalai.
\newblock {Using Large Language Models to Simulate Multiple Humans and
  Replicate Human Subject Studies}.
\newblock In \emph{Proceedings of the 40th International Conference on Machine
  Learning}, 2023.

\bibitem[Argyle et~al.(2023)Argyle, Busby, Fulda, Gubler, Rytting, and
  Wingate]{argyle2023out}
Lisa~P. Argyle, Ethan~C. Busby, Nancy Fulda, Joshua~R. Gubler, Christopher
  Rytting, and David Wingate.
\newblock {Out of One, Many: Using Language Models to Simulate Human Samples}.
\newblock \emph{Political Analysis}, 31\penalty0 (3):\penalty0 337--351, 2023.

\bibitem[Ashery et~al.(2025)Ashery, Aiello, and
  Baronchelli]{ashery2025emergent}
Ariel~Flint Ashery, Luca~Maria Aiello, and Andrea Baronchelli.
\newblock {Emergent Social Conventions and Collective Bias in LLM Populations}.
\newblock \emph{Science Advances}, 11\penalty0 (20):\penalty0 eadu9368, 2025.
\newblock \doi{10.1126/sciadv.adu9368}.

\bibitem[Barab{\'a}si \& Albert(1999)Barab{\'a}si and
  Albert]{barabasi1999emergence}
Albert-L{\'a}szl{\'o} Barab{\'a}si and R{\'e}ka Albert.
\newblock {Emergence of Scaling in Random Networks}.
\newblock \emph{Science}, 286\penalty0 (5439):\penalty0 509--512, 1999.
\newblock \doi{10.1126/science.286.5439.509}.

\bibitem[Burt(2003)]{burt1992structural}
Ronald~S Burt.
\newblock The social structure of competition.
\newblock \emph{Networks in the knowledge economy}, 13\penalty0 (2):\penalty0
  57--91, 2003.

\bibitem[Chen et~al.(2024)Chen, Su, Zuo, Yang, Yuan, Chan, Yu, Lu, Hung, Qian,
  et~al.]{chen2024agentverse}
Weize Chen, Yusheng Su, Jingwei Zuo, Cheng Yang, Chenfei Yuan, Chi-Min Chan,
  Heyang Yu, Yaxi Lu, Yi-Hsin Hung, Chen Qian, et~al.
\newblock Agentverse: Facilitating multi-agent collaboration and exploring
  emergent behaviors.
\newblock In \emph{International Conference on Learning Representations},
  volume 2024, pp.\  20094--20136, 2024.

\bibitem[Cotter et~al.(2001)Cotter, Hermsen, Ovadia, and
  Vanneman]{cotter2001glass}
David~A. Cotter, Joan~M. Hermsen, Seth Ovadia, and Reeve Vanneman.
\newblock {The Glass Ceiling Effect}.
\newblock \emph{Social Forces}, 80\penalty0 (2):\penalty0 655--681, 2001.
\newblock \doi{10.1353/sof.2001.0091}.

\bibitem[Du et~al.(2024)Du, Li, Torralba, Tenenbaum, and
  Mordatch]{du2024multiagentdebate}
Yilun Du, Shuang Li, Antonio Torralba, Joshua~B. Tenenbaum, and Igor Mordatch.
\newblock Improving factuality and reasoning in language models through
  multiagent debate.
\newblock In Ruslan Salakhutdinov, Zico Kolter, Katherine Heller, Adrian
  Weller, Nuria Oliver, Jonathan Scarlett, and Felix Berkenkamp (eds.),
  \emph{Proceedings of the 41st International Conference on Machine Learning},
  volume 235 of \emph{Proceedings of Machine Learning Research}, pp.\
  11733--11763. PMLR, 21--27 Jul 2024.
\newblock URL \url{https://proceedings.mlr.press/v235/du24e.html}.

\bibitem[Gao et~al.(2023)Gao, Lan, Lu, Mao, Piao, Wang, Jin, and Li]{gao2023s3}
Chen Gao, Xiaochong Lan, Zhihong Lu, Jinzhu Mao, Jinghua Piao, Huandong Wang,
  Depeng Jin, and Yong Li.
\newblock S3: Social-network simulation system with large language
  model-empowered agents.
\newblock \emph{arXiv preprint arXiv:2307.14984}, 2023.

\bibitem[Granovetter(1973)]{granovetter1973strength}
Mark~S Granovetter.
\newblock The strength of weak ties.
\newblock \emph{American journal of sociology}, 78\penalty0 (6):\penalty0
  1360--1380, 1973.

\bibitem[Guan et~al.(2025)Guan, He, Fan, Ren, He, Yu, Chen, Zheng, Liu, and
  Liu]{guan2025modeling}
Haoxiang Guan, Jiyan He, Liyang Fan, Zhenzhen Ren, Shaobin He, Xin Yu, Yuan
  Chen, Shuxin Zheng, Tie-Yan Liu, and Zhen Liu.
\newblock Modeling earth-scale human-like societies with one billion agents.
\newblock \emph{arXiv preprint arXiv:2506.12078}, 2025.

\bibitem[Guo et~al.(2026)Guo, Wu, and Yiu]{guo2026coalition}
Dongxin Guo, Jikun Wu, and Siu-Ming Yiu.
\newblock Coalition formation in llm agent networks: Stability analysis and
  convergence guarantees.
\newblock \emph{arXiv preprint arXiv:2604.14386}, 2026.

\bibitem[Guo et~al.(2024)Guo, Chen, Wang, Chang, Pei, Chawla, Wiest, and
  Zhang]{guo2024large}
Taicheng Guo, Xiuying Chen, Yaqi Wang, Ruidi Chang, Shichao Pei, Nitesh~V.
  Chawla, Olaf Wiest, and Xiangliang Zhang.
\newblock {Large Language Model Based Multi-Agents: A Survey of Progress and
  Challenges}.
\newblock \emph{arXiv preprint arXiv:2402.01680}, 2024.

\bibitem[Hong et~al.(2024)Hong, Zhuge, Chen, Zheng, Cheng, Wang, Zhang, Yau,
  Lin, Zhou, et~al.]{hong2024metagpt}
Sirui Hong, Mingchen Zhuge, Jonathan Chen, Xiawu Zheng, Yuheng Cheng, Jinlin
  Wang, Ceyao Zhang, Steven Yau, Zijuan Lin, Liyang Zhou, et~al.
\newblock Metagpt: Meta programming for a multi-agent collaborative framework.
\newblock In \emph{International Conference on Learning Representations},
  volume 2024, pp.\  23247--23275, 2024.

\bibitem[Horton et~al.(2023)Horton, Filippas, and Manning]{horton2023large}
John~J Horton, Apostolos Filippas, and Benjamin~S Manning.
\newblock Large language models as simulated economic agents: What can we learn
  from homo silicus?
\newblock Technical report, National Bureau of Economic Research, 2023.

\bibitem[Hu et~al.(2026)Hu, Tan, Wang, Qu, and Chen]{hu2025multiagent}
Tianyu Hu, Zhen Tan, Song Wang, Huaizhi Qu, and Tianlong Chen.
\newblock Multi-agent debate for llm judges with adaptive stability detection.
\newblock \emph{Advances in Neural Information Processing Systems},
  38:\penalty0 46504--46540, 2026.

\bibitem[Jain \& Krishnamurthy(2024)Jain and
  Krishnamurthy]{jain2024interacting}
Adit Jain and Vikram Krishnamurthy.
\newblock {Interacting Large Language Model Agents. Interpretable Models and
  Social Learning}.
\newblock \emph{arXiv preprint arXiv:2411.01271}, 2024.

\bibitem[Jain et~al.(2025{\natexlab{a}})Jain, Krishnamurthy, and
  Zhang]{jain2025collaborative}
Adit Jain, Vikram Krishnamurthy, and Yiming Zhang.
\newblock {Collaborative QA using Interacting LLMs. Impact of Network
  Structure, Node Capability and Distributed Data}.
\newblock \emph{arXiv preprint arXiv:2511.14098}, 2025{\natexlab{a}}.

\bibitem[Jain et~al.(2025{\natexlab{b}})Jain, Krishnamurthy, and
  Zhang]{jain2025information}
Adit Jain, Vikram Krishnamurthy, and Yiming Zhang.
\newblock {Information Diffusion and Preferential Attachment in a Network of
  Large Language Models}.
\newblock In \emph{2025 IEEE 64th Conference on Decision and Control (CDC)},
  pp.\  180--185, 2025{\natexlab{b}}.
\newblock \doi{10.1109/CDC57313.2025.11312386}.

\bibitem[Kushner \& Yin(2003)Kushner and Yin]{KushnerYin2003}
Harold~J. Kushner and George~G. Yin.
\newblock \emph{{Stochastic Approximation and Recursive Algorithms and
  Applications}}, volume~35 of \emph{Applications of Mathematics}.
\newblock Springer, New York, 2 edition, 2003.

\bibitem[Li et~al.(2023)Li, Hammoud, Itani, Khizbullin, and
  Ghanem]{li2023camel}
Guohao Li, Hasan Hammoud, Hani Itani, Dmitrii Khizbullin, and Bernard Ghanem.
\newblock Camel: Communicative agents for" mind" exploration of large language
  model society.
\newblock \emph{Advances in neural information processing systems},
  36:\penalty0 51991--52008, 2023.

\bibitem[Li et~al.(2024)Li, Du, Zhang, Hou, Grabowski, Li, and
  Ie]{li2024sparse}
Yunxuan Li, Yibing Du, Jiageng Zhang, Le~Hou, Peter Grabowski, Yeqing Li, and
  Eugene Ie.
\newblock {Improving Multi-Agent Debate with Sparse Communication Topology}.
\newblock In \emph{Findings of the Association for Computational Linguistics:
  EMNLP 2024}, pp.\  7281--7294, 2024.
\newblock URL \url{https://aclanthology.org/2024.findings-emnlp.427/}.

\bibitem[Luo et~al.(2024)Luo, Nettasinghe, and Krishnamurthy]{luo2024mutual}
Rui Luo, Buddhika Nettasinghe, and Vikram Krishnamurthy.
\newblock {Mutual Information Measure for Glass Ceiling Effect in Preferential
  Attachment Models}.
\newblock \emph{IEEE Transactions on Computational Social Systems}, 11\penalty0
  (6):\penalty0 7778--7788, 2024.

\bibitem[Madmoun \& Lahlou(2026)Madmoun and Lahlou]{madmoun2025communication}
Hachem Madmoun and Salem Lahlou.
\newblock Communication enables cooperation in llm agents: A comparison with
  curriculum-based approaches.
\newblock In \emph{Proceedings of the 19th Conference of the European Chapter
  of the Association for Computational Linguistics (Volume 2: Short Papers)},
  pp.\  307--321, 2026.

\bibitem[Mehdizadeh \& Hilbert(2025)Mehdizadeh and
  Hilbert]{mehdizadeh2025homophily}
Aliakbar Mehdizadeh and Martin Hilbert.
\newblock Homophily-induced emergence of biased structures in llm-based
  multi-agent ai systems.
\newblock \emph{Social Network Analysis and Mining}, 15\penalty0 (1):\penalty0
  1--25, 2025.

\bibitem[Merton(1968)]{merton1968matthew}
Robert~K Merton.
\newblock The matthew effect in science: The reward and communication systems
  of science are considered.
\newblock \emph{Science}, 159\penalty0 (3810):\penalty0 56--63, 1968.

\bibitem[Nettasinghe et~al.(2022)Nettasinghe, Alipourfard, Iota, Krishnamurthy,
  and Lerman]{nettasinghe2022scale}
Buddhika Nettasinghe, Nazanin Alipourfard, Stephen Iota, Vikram Krishnamurthy,
  and Kristina Lerman.
\newblock Scale-free degree distributions, homophily and the glass ceiling
  effect in directed networks.
\newblock \emph{Journal of complex networks}, 10\penalty0 (2):\penalty0
  cnac007, 2022.

\bibitem[Nettasinghe et~al.(2026)Nettasinghe, Alipourfard, Krishnamurthy, and
  Lerman]{nettasinghe2026emergence}
Buddhika Nettasinghe, Nazanin Alipourfard, Vikram Krishnamurthy, and Kristina
  Lerman.
\newblock Emergence of structural disparities in the web of scientific
  citations.
\newblock In \emph{Proceedings of the ACM Web Conference 2026}, pp.\
  1785--1796, 2026.

\bibitem[Papachristou \& Yuan(2025)Papachristou and
  Yuan]{papachristou2025network}
Marios Papachristou and Yuan Yuan.
\newblock Network formation and dynamics among multi-llms.
\newblock \emph{PNAS nexus}, 4\penalty0 (12):\penalty0 pgaf317, 2025.

\bibitem[Park et~al.(2023)Park, O'Brien, Cai, Morris, Liang, and
  Bernstein]{park2023generative}
Joon~Sung Park, Joseph O'Brien, Carrie~Jun Cai, Meredith~Ringel Morris, Percy
  Liang, and Michael~S Bernstein.
\newblock Generative agents: Interactive simulacra of human behavior.
\newblock In \emph{Proceedings of the 36th annual acm symposium on user
  interface software and technology}, pp.\  1--22, 2023.

\bibitem[Piao et~al.(2026)Piao, Yan, Zhang, Li, Yan, Lan, Lu, Zheng, Wang,
  Zhou, Gao, Xu, Zhang, Rong, Su, and Li]{piao2025agentsociety}
Jinghua Piao, Yuwei Yan, Jun Zhang, Nian Li, Junbo Yan, Xiaochong Lan, Zhihong
  Lu, Zhiheng Zheng, Jing~Yi Wang, Di~Zhou, Chen Gao, Fengli Xu, Fang Zhang,
  Ke~Rong, Jun Su, and Yong Li.
\newblock Agentsociety: Large-scale simulation of llm-driven generative agents
  advances understanding of human behaviors and society, 2026.
\newblock URL \url{https://arxiv.org/abs/2502.08691}.

\bibitem[Price(1976)]{price1976general}
Derek De~Solla Price.
\newblock A general theory of bibliometric and other cumulative advantage
  processes.
\newblock \emph{Journal of the American society for Information science},
  27\penalty0 (5):\penalty0 292--306, 1976.

\bibitem[Qian et~al.(2024)Qian, Liu, Liu, Chen, Dang, Li, Yang, Chen, Su, Cong,
  et~al.]{qian2024chatdev}
Chen Qian, Wei Liu, Hongzhang Liu, Nuo Chen, Yufan Dang, Jiahao Li, Cheng Yang,
  Weize Chen, Yusheng Su, Xin Cong, et~al.
\newblock Chatdev: Communicative agents for software development.
\newblock In \emph{Proceedings of the 62nd annual meeting of the association
  for computational linguistics (volume 1: Long papers)}, pp.\  15174--15186,
  2024.

\bibitem[Qian et~al.(2025)Qian, Xie, Wang, Liu, Zhu, Xia, Dang, Du, Chen, Yang,
  et~al.]{qian2024scaling}
Chen Qian, Zihao Xie, Yifei Wang, Wei Liu, Kunlun Zhu, Hanchen Xia, Yufan Dang,
  Zhuoyun Du, Weize Chen, Cheng Yang, et~al.
\newblock Scaling large language model-based multi-agent collaboration.
\newblock In \emph{International Conference on Learning Representations},
  volume 2025, pp.\  41488--41505, 2025.

\bibitem[Schneider et~al.(2025)Schneider, Tian, and
  Rizoiu]{schneider2025learning}
Philipp~J Schneider, Lin Tian, and Marian-Andrei Rizoiu.
\newblock Learning to make friends: Coaching llm agents toward emergent social
  ties.
\newblock \emph{arXiv preprint arXiv:2510.19299}, 2025.

\bibitem[Vaswani et~al.(2017)Vaswani, Shazeer, Parmar, Uszkoreit, Jones, Gomez,
  Kaiser, and Polosukhin]{vaswani2017attention}
Ashish Vaswani, Noam Shazeer, Niki Parmar, Jakob Uszkoreit, Llion Jones,
  Aidan~N Gomez, {\L}ukasz Kaiser, and Illia Polosukhin.
\newblock {Attention is all you need}.
\newblock \emph{Advances in neural information processing systems}, 30, 2017.

\bibitem[Wu et~al.(2023)Wu, Bansal, Zhang, Wu, Li, Zhu, Jiang, Zhang, Zhang,
  Liu, et~al.]{wu2023autogen}
Qingyun Wu, Gagan Bansal, Jieyu Zhang, Yiran Wu, Beibin Li, Erkang Zhu,
  Li~Jiang, Xiaoyun Zhang, Shaokun Zhang, Jiale Liu, et~al.
\newblock Autogen: Enabling next-gen llm applications via multi-agent
  conversation.
\newblock \emph{arXiv preprint arXiv:2308.08155}, 2023.

\end{thebibliography}
\end{document}